\documentclass[12pt]{article}
\usepackage{graphicx}
\small

\begin{document}

\centerline {\bf ROTATION RATE OF HIGH LATITUDE AND NEAR POLAR CORONAL HOLES}
\vskip 0.3cm
\centerline {\em K. M. Hiremath$^{1,2}$, Manjunath Hegde$^{1}$ and K. R. Varsha$^{3}$}

1. Indian Institute of Astrophysics, Bangalore, India, Email: hiremath@iiap.res.in

2. \#23, Mathru Pithru Krupa, 2nd Cross, 1st Main, BDA Layout, Bikasipura, BSK V Stage, Bengaluru-560111, India

3. No 204, 2nd Cross, VBHCS, R.R Nagar, Bengaluru - 560098, India

\vskip 0.3cm

\centerline {\bf ABSTRACT}
For the period of 1997-2006, coronal holes detected
in the SOHO/EIT 195 $\AA$ full disk
calibrated images are used to compute the rotation rates
of high latitude and near polar coronal holes and, their latitudinal variation is
investigated. We find that, for different latitude zones between
$80^{o}$ north and $75^{o}$ south,  for all their area,
the number of days observed on the solar disk, and their latitudes,
coronal holes rotate rigidly. Estimated magnitudes of sidereal rotation rate
of the coronal holes are: $13.051 \pm 0.206$ deg/day for the equator,
 $12.993 \pm 0.064$ deg/day in the region of higher latitudes and, $12.999 \pm 0.329$ deg/day
near the polar regions.  
 For all the latitudes
and the area, we have also investigated
the annual variation of rotation rates of these coronal holes.
We find that, for all the years, coronal holes
rotate rigidly and their magnitude of equatorial, high latitude and polar region rotation rates
 are independent of magnitude of solar activity.

\section{INTRODUCTION}

 Solar coronal holes (CHs) are observed in EUV and Xray images as dark features, whereas these activity features appear  as bright
features in the  He I 10830 $\AA$ (Harvey and Sheeley 1979). Depending upon observed wavelengths, coronal holes either
partly or predominantly are associated with the large-scale unipolar magnetic field
structure of the sun. CHs are large regions in the solar corona
with low density plasma (Krieger {\em et. al} 1973, Neupert and Pizzo 1974, Nolte {\em et.al.} 1976,
Zirker 1977, Cranmer 2009, Wang 2009).
CHs are the source regions of fast solar wind (Altschuler 1972, Temmer {\em et. al.} 2007, Hegde {\em et.al.} 2015).
Recent studies (Soon {\em et.al.} 2000, Shugai {\em et.al.} 2009, Verbanac {\em et.al.} 2011)
show that (Harvey and Sheeley 1979, Krieger {\em et. al} 1973, Neupert and Pizzo 1974, Nolte {\em et.al.} 1976, Zirker 1977)
on short time scales, occurrences of solar coronal holes trigger
responses in the Earth's upper atmosphere and magnetosphere
apart from sunspot and magnetic activity (Hiremath 2009 and references there in).

 For better understanding of solar activity, idea of rotational structure of
different layers i.e, interior, surface and atmosphere of sun are essential. Currently,
there is a general consensus regarding the interior rotation as inferred
from the helioseismology (Dalsgaard and Schou, Antia {\em et.al.} 1998, Thompson {\em et.al} 2003, 
Howe 2009, Antia and Basu 2010). Also, we have the knowledge of
surface rotation rates derived from sunspots
(Newton and Nunn 1951, Howard {\em et.al.} 2084, Balthasar {\em et.al} 1986, Sivaraman {\em et.al.} 1993, Javaraiah 2003),
Doppler velocity (Howard and Harvey 1970, Ulrich {\em et.al} 1988, Snodgrass and Ulrich 1990) and magnetic activity features
(Wilcox and Howard 1970, Snodgrass 1983, Komm {\em et.al.} 1993). But the
 magnitude and form of rotation law for features in the corona show different
natures (Hiremath and Hegde 2013, and references there in).

Although there is a general consensus regarding surface (derived from the sunspots and
Doppler velocity) and internal rotation (inferred from the helioseismology) rate
of the sun, whereas right from discovery of coronal holes, there is no consensus
regarding rotation rate, whether these coronal activities rotate rigidly or differentially.
From the detected coronal holes in different spectral windows, some studies
(Shelke and Pande 1985, Obridko and Shelting 1989, Navaro and Sanchez 1994, Insley {\em et. al.} 1995, Oghrapishvili {\em et. al.} 2018) show
that coronal holes rotate differentially and, other majority of studies came to a conclusion  that the
coronal holes rotate rigidly (Wagner 1975, Wagner 1976, Timothy {\em et. al.} 1975, Bohlin 1977, 
 Hiremath and Hegde 2013, Japaridze {\em et.al.} 2015, Bagashvili {\em et.al} 2017, Prabhu {\em et.al} 2018).

 Using KPNO helium synoptic charts, rotation rates of equatorial and high latitude
coronal holes (Shelke and Pande 1985) are computed and it is found that coronal
holes rotate rigidly. For the years 1978-1986, Obridko and Shelting (1989) analyzed the synoptic charts
of solar geophysical data and came to the conclusion that coronal holes rotate differentially. 
From the analysis of of 18 yrs of coronal hole data, Navaro and Sanchez (1994) came to the
following conclusions: (i) for all latitude, coronal holes rotate differentially,
(ii) below 40 deg latitude, magnitude of rotation rate of coronal holes increases
towards the equator, (iii) whereas, rotation rate of coronal holes above 40 deg latitude
increases towards pole. From the analysis of He 10830 $\AA$ coronal hole data, 
Iinsle {\em et. al.} (1995) found that coronal holes rotate differentially. Very recently,
Oghrapishvili (2018) used two years SDO/AIA data of coronal holes and similar
to previous studies, these authors also came to a conclusion that coronal holes
rotate differentially.

 Pioneering studies (Wagner 1975, Wagne 1976) on the rotation rate of the coronal
holes that are analyzed from the Fe XV (284 \AA) data showed that 
coronal holes rotate rigidly. The coronal holes observed in the same wave length
were used by Timothy {\em et. al.} (1975) and obtained the result that coronal holes
rotate rigidly. From the evolutionary and rotational characteristics, these authors
conclude that coronal holes probably might have originated below the photosphere.
Continuing with the same coronal hole data, Bohlin (1977) also came to a similar
conclusion of rigid body rotation rate of the coronal holes. For the years
2001-2008 and from the full-disk SOHO/EIT 195 $\AA$ near equatorial coronal hole data, Hiremath and Hegde (2013)
showed that irrespective their area and latitude, coronal holes rotate rigidly.
Further, from the area evolutionary and dynamical characteristics, with reasonable
theoretical arguments, Hiremath and Hegde (2013) showed that coronal holes probably
might be originated below base of the convection zone.  From the Kitt Peak Observatory 
He I 10830 $\AA$ coronal hole data and for the years 2003-2012, Japaridze {\em et.al.} 2015 
showed that coronal holes have magnitude of rotation rate of 13.39 deg/day and rotate
rigidly compared to photospheric tracer rotation rates. Bagashvili {\em et.al} (2017) analyzed
SDO/AIA 193 $\AA$ coronal hole data and two important results relevant to the present study
are: (i) coronal holes rotate rigidly and, (ii) coronal hole rotation rate
latitudinal profile perfectly matches only with the latitudinal rotational
rate of the solar plasma around 0.71 $R_{\odot}$, near lower layers of base 
of convection zone. Very recent study (Prabhu {\em et.al} 2018) that used the
recurrent coronal hole data of He 10830 $\AA$ and EUV 195 $\AA$ came to the conclusion
that coronal holes rotate rigidly and conjectured that these coronal hole
activity features might have originated below base of convection zone.
  
Most probable reasons for these contradicting results on rotation rate of the coronal
holes are the following: (i) lack of correct methods for detection of coronal holes, especially
during the initial period of discovery of coronal holes, and their
accurate determination of heliographic coordinates and, (ii) not taking into
account the projectional effects, especially near higher latitudes and near the limbs, for the position of coronal holes.
Infact, we have fulfilled these two demands reasonably and accurately in the
present study. It is also to be noted that most of the afore mentioned studies
restricted with in the latitudes of 60 deg north to 60 deg south.
However, it is interesting to know whether coronal holes near the
polar regions rotate rigidly or differentially.

In addition, there are also following two studies which conclude that, in the lower latitudes, coronal
holes rotate differentially whereas in the higher latitudes these features
rotate rigidly.  Based on the Catalog of Coronal Holes, Navaro and Sanchez (1994)
 found that isolated or equatorial
CHs show differential rotation rate while the polar CHs
 exhibit rigid body rotation rate. Zurbuchen {\em et.al.} (1996) using SWICS/Ulysses data
found that, in the near equatorial and intermediate latitudes, a quasi-rigid rotation rate and no persistent structures
at latitudes \textgreater 65$^\circ$  to determine the polar rotation rate. In continuation of our
previous study (Hiremath and Hegde 2013), in the present study,
we extend the investigation of rotation rates of coronal holes from near equator
to higher latitudes and near polar regions. Although definitions are ad-hoc, according to our
definition, equatorial coronal holes are confined to +45 to -45 deg, the higher latitude
coronal holes occur in the region of 46 to 65 deg northern and southern hemispheres
and, the near polar coronal holes originate in the latitude belt of 66-90 deg in northern and southern hemispheres.

Compared to our previous study (Hiremath and Hegde 2013), this
study is different on the four important aspects: (i) in addition to near equatorial coronal holes, in the present study,
high latitude and near polar coronal hole rotation rates are also estimated, (ii) in the previous study, we used to detect manually the coronal holes, either ellipse or circle is fitted to the coronal hole and average heliographic coordinates
and the area are used to be estimated. However, in the present study, from the morphological image analysis, coronal hole boundary
is detected automatically and, average heliographic coordinates and the areas are estimated consistently without any bias, (iii) in the
previous study, correct formula for estimation of projectional effects of coronal holes is not used, whereas in the present
study correct formula is used, (iv) in the present study, yearly change in rotation rate of coronal holes is investigated, whereas
in the previous study it is not. 

Using photospheric sunspot and other activity features, there are many studies that
deal with the variation of rotation rate of the sun with respect to solar cycle
activity. In the present study we also examine variation of rotation rate of coronal
holes and different coefficients of law of rotation rates with respect to activity
of the solar cycle. Although it appears to be of similar results for all the latitudes, question remains from
the previous study that whether coronal holes (at the high-latitude and near polar regions) rotate rigidly or differentially; even if it is
either rigid or differential body rotation, it is interesting to know sign of third coefficient in the of rotation law (obtained from
the least-square fit) is positive or negative (please to be noted that, according to Snodgrass, sign of the third coefficient is negative,
whereas interestingly for the coronal holes it is positive, at present we don't know why this sign is positive).

In section 2, we present the data used and method of analysis, and results of
that analysis in section 3. While section 4 ends with the conclusions of the present study.

 \section{DATA AND ANALYSIS}
 From 1997-2006, we use full-disk SOHO/EIT images (Delaboudiniere 1995) that have a resolution of 2.6 arc sec per pixel in a band pass around 
195 $\AA$ to detect CHs. It is to be noted that there are data gaps in SOHO for June-October 1998 and January-February 1999.
 The obtained images are in FITS format and the individual pixels are in units of data number (DN). DN is defined as the output of the instrument electronics that corresponds to the incident photon signal converted into charge within each CCD pixel (Madjarska and Wiegelmann 2009).

 Using the SolarSoft eit\_prep routine (Freeland and Handy 1998),
images are  background subtracted, flat-fielded, degridded, and
are normalized. Occurrence dates and position of CH
are obtained from the ``{\em spaceweather.com''} web site.
By using the approximate positions (heliographic coordinates) of
CHs from the ``{\em spaceweather.com''} web site, coronal hole region from the SOHO/EIT images
are separated for further analysis.
 We use intensity thresholding technique to identify coronal holes in the images.
 Krista and Gallagher (2009) pointed out that in case of coronal hole
the intensity histogram shows a bimodal distribution (see the middle illustration of intensity histogram in both the upper and lower
panels of Fig 1) and the local minimum corresponds to
CH boundary threshold. Following this important fact and for each daily image we find this local intensity minimum that is used as a threshold to detect
coronal holes in all the latitudes. So, regardless of the phases of solar cycle this method works very well for all the images.

From the basic morphological image processing operations like erosion and dilation, boundary of the coronal holes
are detected. With the SolarSoft, individual pixel information (heliographic
coordinates such as latitude and longitude and, the DN counts/intensity) enclosing the coronal holes is obtained.
 Following our previous study (Hiremath and Hegde 2013, equation 1), average heliographic coordinates weighted with intensity
(DN counts) is obtained. As each pixel is weighted with intensity, a reasonable average
of these two important parameters can be obtained. With this method, errors in the average coordinates
can also be obtained. In this way accuracy of these two coordinates can also be estimated.

   We followed the following criteria in selecting CHs data.
 (1) the CH must be compact, independent, and is not elongated, for example, from pole to
the equator in latitude ,
and (2) in order to maintain the original identity, consider only CHs  that  do not merge
with other CHs during their passage. Although this data consists of both low and high
latitude data, as for high latitude coronal
holes, for each pixel of the detected coronal hole,
following corrections for the projectional effects are applied
for intensity $I$ (DN counts)

\begin{equation} 
        I = {I_{obs} \over{cos\delta}} \, ,
\end{equation}
\noindent where, $I_{obs}$ is the observed intensity of the coronal hole,
$cos\delta=(sinB_{0} sin \theta+cos B_{0} cos \theta cos l$),
$\theta$ and $l$ are heliographic latitude and heliographic
longitude from the central meridian of the CH respectively.
This projectional ($cos\delta$ term in the denominator of eqn 1) correction for the intensity of pixel takes
into account high latitude and extreme longitude (from the central meridian) coronal holes.
Whereas $B_{0}$ is the heliographic latitude of the center of the
solar disk at the time of observation. After applying correction for projectional effects and following
Hiremath and Hegde (2013) (section 2, eqn 1), average heliographic latitude and longitude
from the central meridian of the coronal hole are estimated.

Following the previous method (Hiremath 2002) of computation of
rotation rates of sunspots, daily sidereal rotation rates $\Omega_{j}$ of the CH are computed as follows
\begin{equation}
\Omega_{j}=\frac{(L_{j+1} - L_{j})}{(t_{j+1} - t_{j})} + \delta \Omega \, ,
\end{equation}
where L$_{j}$, L$_{j+1}$ are average longitude from the central meridian of the CH for the two consecutive days
 t$_{j}$ and t$_{j+1}$ respectively, $j=1,2, ..n-1$, $n$ is number
of days of appearance of CH on the visible solar disk and,
 $\delta \Omega$ is a correction factor for the
orbital motion of the Earth around the sun.
Strictly speaking, this correction factor is due to orbital motion of the SOHO spacecraft around the sun.
 Compared to the distance between the sun and Earth, the
distance between the SOHO satellite and the Earth is very small and hence
orbital distances of Earth and the satellite are almost same and hence
the correction factor $\delta \Omega$ is $\sim$ 1 deg/day in order
to get the sidereal rotation rate of the coronal holes. For the present work, this approximation
is sufficient. However, if one wants to find the long term (greater than 11 yrs)
variation of rotation rates, correction factor $\delta \Omega$
should be computed accurately (Ro\v{s}a {\em et.al.} 1995, Wittmann 1996, Skokic {\em et.al.} 2014).

\section{RESULTS}

For the period of observations from 1997 to 2006, a total of
163 CHs satisfy the criteria as given in the previous section.
For different years such occurrence  number of coronal
holes is presented in Fig 2 (a).
During their evolutionary passage over the solar disk, we compute
rotation rates and assign respective latitudes. 
In the present study, we consider only non-recurrent CHs that appear and disappear on same part of the visible disk.
 If such a non-recurrent CH exists for $n$ days, then its life span $\tau$ is $n$ days and the total number of rotation rates is $(n - 1)$.  
 Irrespective of their area, in both the hemispheres,
rotation rates of coronal holes for
each $5^{o}$ bin are  collected and, average rotation rate and its standard deviation are computed.
For example, Fig 2(b) illustrates the idea of number of rotation rates in
each bin considered for computation of average rotation rate.

\subsection{Latitudinal variation of coronal holes}

With the constraints of 65 deg east to 65 west of longitude from
central meridian and for latitude zone of +45 deg to -45 deg, Figure 3(a)
illustrates the latitudinal variation of rotation rate of coronal
holes. In this and subsequent Figures, units of rotation rate in the left hand side of $y$ axis
is deg/day and on the right hand side of $y$ axis is nHz (in order
to compare with the rotation rate of sun's interior inferred by the
helioseismology which use nHz as the unit). 

 With a similar constraint of latitude (65 deg east to 65 west) zone, 
for the same SOHO/EIT 195 $\AA$ and for the years 2001-2008, in the previous study (Hiremath and Hegde 2013, Fig 4(a)) 
we have obtained the rotation rate profile that yields coronal holes 
rotate differentially. In fact, it is to be noted that, in terms of magnitude (of both the coefficients),
 differential rotation rate profile obtained in the present study is almost similar
to the rotation rate profile as obtained by our previous study.
Whereas, for the same constraint on longitudes,
 Fig 3(b) illustrates the variation of rotation rate of the coronal holes
for latitude zone that varies from 80 deg north to 75 deg south.
 That means the results presented in both
the Figures 3(a) and 3(b) suggest that for all the latitudes
coronal holes rotate differentially. In the following, let us examine further whether
this picture changes in case we constrain further in longitudes
and after taking into account the projectional effects.

For the same constraint of latitude zone, albeit with
a constraint of 45 deg east to 45 west longitude zone
(from the central meridian), Figures 4(a) and 4(b) illustrate
the latitudinal variation of rotation rate of  the coronal
holes. However, in addition to mentioned constraints, in case of Fig 4(b),
projectional effect is also taken into account. Note that
when we apply correct constraints, picture of differential
rotation rate of the coronal holes changes to rigid body
rotation rate. Infact, this conclusion can also be judged
from the consistent decrease of magnitude of second coefficient
(that represents differential rotation of the  mid-latitudes)
of law obtained from the least square fit and also
decrease in the value of $\chi^{2}$ (a measure of
goodness of fit). Of course one can also argue that the law of
fit is up to $sin^{2} \theta$ which may not correctly represents the
data of high latitude and near polar region coronal holes, for
which law of fit should be up to $sin^{4} \theta$. In the
following, law of fit up to $sin^{4} \theta$ is used and examined
whether coronal holes rotate differentially or rigidly.

For the constraint of 65 deg east to 65 deg west of longitude
from the central meridian, Fig 5(a) illustrates the latitudinal
variation of rotation rate of coronal holes whose profile
is fitted with a law up to $sin^{4} \theta$. Whereas,
with the constraint of 45 deg east to 45 deg west
of longitudes, Fig 5(b) illustrates the variation of rotation rate
of coronal holes whose profile is also fitted with
the rotational law up to $sin^{4} \theta$. Although both the
results illustrated in Fig 5(a) and Fig 5(b) appear that
coronal holes rotate differentially, when we apply the
projectional corrections, the results unambiguously
suggest that  (as illustrated in Fig 6) for all the
latitudes coronal holes rotate rigidly.

If we compare the
coefficients (the law over plotted on Fig 6)
of $sin^{4} \theta$ obtained by the least square fit
with the coefficients of $sin^{4} \theta$ obtained by
other study (Snodgrass 1983 and references there in)
which uses high latitude magnetic activity features and
surface plasma rotation (by Doppler shift) as tracers,
obviously for all the latitudes coronal holes rotate
rigidly. For further clarity of this statement, let us
compare both the laws. Snodgrass laws for both the tracers
yield: magnetic tracers: $\Omega (\theta) = (14.3065 - 1.9801 sin^{2} \theta - 2.1485 sin^{4} \theta) \,  deg/day$ ; Doppler velocity: $\Omega (\theta) = (14.1432 - 1.8132  sin^{2} \theta - 2.4925  sin^{4} \theta) \,  deg/day$. When we compare Snodgrass surface rotation
rate laws with the rotation rate law ($\Omega (\theta) = 13.051-0.161 sin^{2} \theta + 0.111 sin^{4} \theta$) obtained
from coronal holes,  following are the notable differences:
 (i) magnitude of first coefficient  is lesser by $\sim$ 7.2$\%$ compared to magnitude
of first coefficient of Snodgrass laws,
(ii) second and third coefficients are nearly 10$\%$  of the
second and third coefficients in the Snodgrass laws and,
(iii) there is a change in sign ( positive) for the coefficient
of $sin^{4} \theta$ compared to the sign (negative) of
coefficient of $sin^{4} \theta$ in case of Snodgras laws. Hence, overall conclusion of this
subsection is that, {\em for all the area and latitude, coronal
holes rotate rigidly}.

Another interesting aspect of this study is that, if we consider the first
coefficient of rotation law (see Fig 6), irrespective of all the latitudes,
coronal holes rotate with a magnitude of $\sim$ 420 nHz. Infact, previous
studies (Hiremath and Hegde 2013, Japaridze {\em et.al.}  2015, Bagashvili {\em et.al.} 2017, Prabhu {\em et.al.} 2018)
came to a conclusion that anchoring depth of coronal holes is probably
below base the convection zone. If we consider magnitude of average rotation rate of coronal hole
which  is estimated to be $\sim$ 420 nHz, this can be matched only with the
rotation rate of the deep interior of radiative core (Antia {\em et.al.} 1998, Antia and Basu 2010).
However, at present, it is not clear how such gigantic coronal holes form
in the deep interior and then protrude into the solar atmosphere, corona and beyond.
As the genesis of coronal hoes is beyond the scope of this study,
further theoretical studies and helioseismic inferences are required
to prove or disprove the proposition that coronal holes originate in the
deep interior.

\begin{figure}
\begin{center}
\vskip -8.5ex
      {\includegraphics[width=40pc,height=30pc]{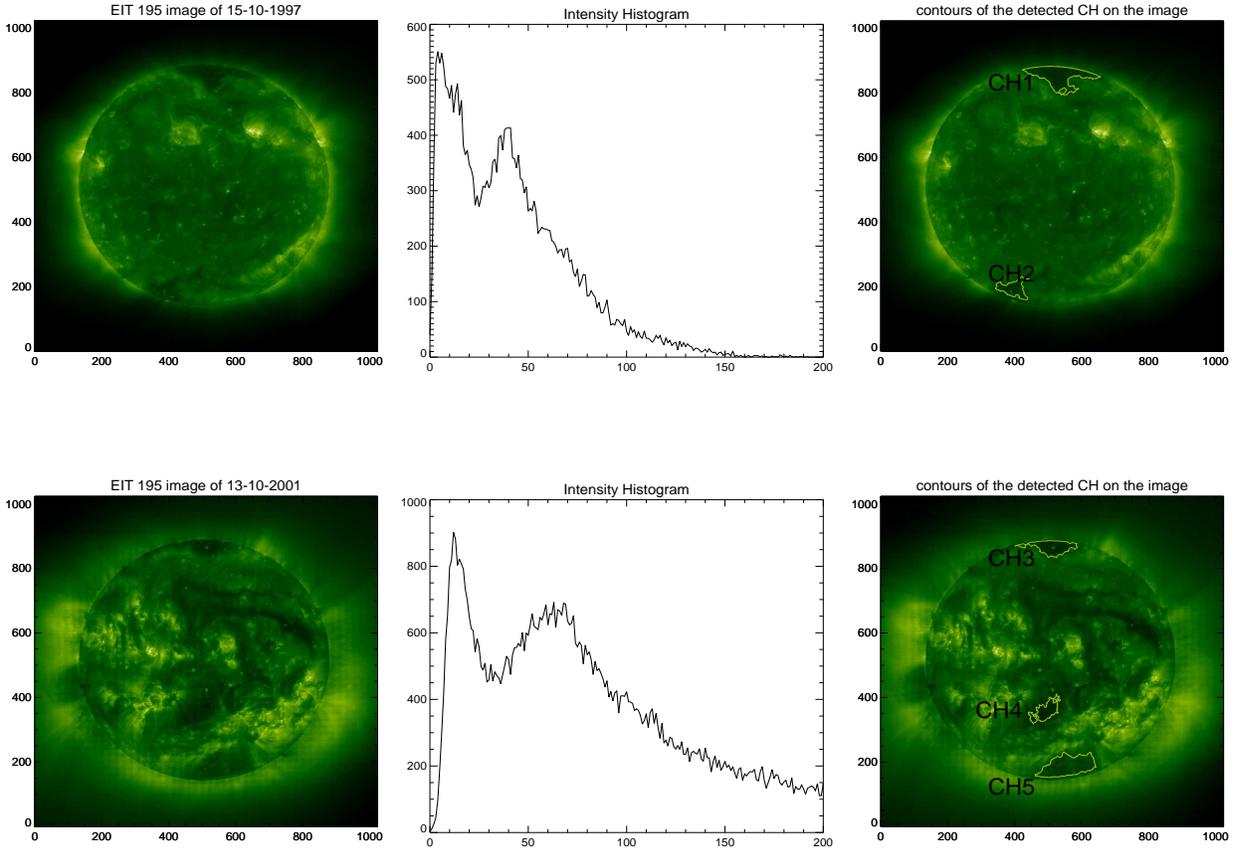}}
    \caption{ For the observed date 15-10-1997, three successive figures in upper panel represent:
SOHO/MDI full disk image observed in 195 $\AA$, intensity histogram of the full disk image and the
detected coronal holes with isocontours as boundary. Whereas lower panel represents for the
date 13-10-2001. In both the panels of images CH1 (latitude 75.03 deg North), CH2 (69.94 deg South),
CH3 (76.85 deg South), CH4 (20.59 South) and CH5(61.05 South) are the detected coronal holes
respectively.   }
\end{center}
\end{figure}

\begin{figure}
\begin{center}
     Fig 2(a) \hskip 25ex  Fig 2(b)
\vskip -4.5ex
    \begin{tabular}{cc}
      {\includegraphics[width=20.0pc,height=20.0pc]{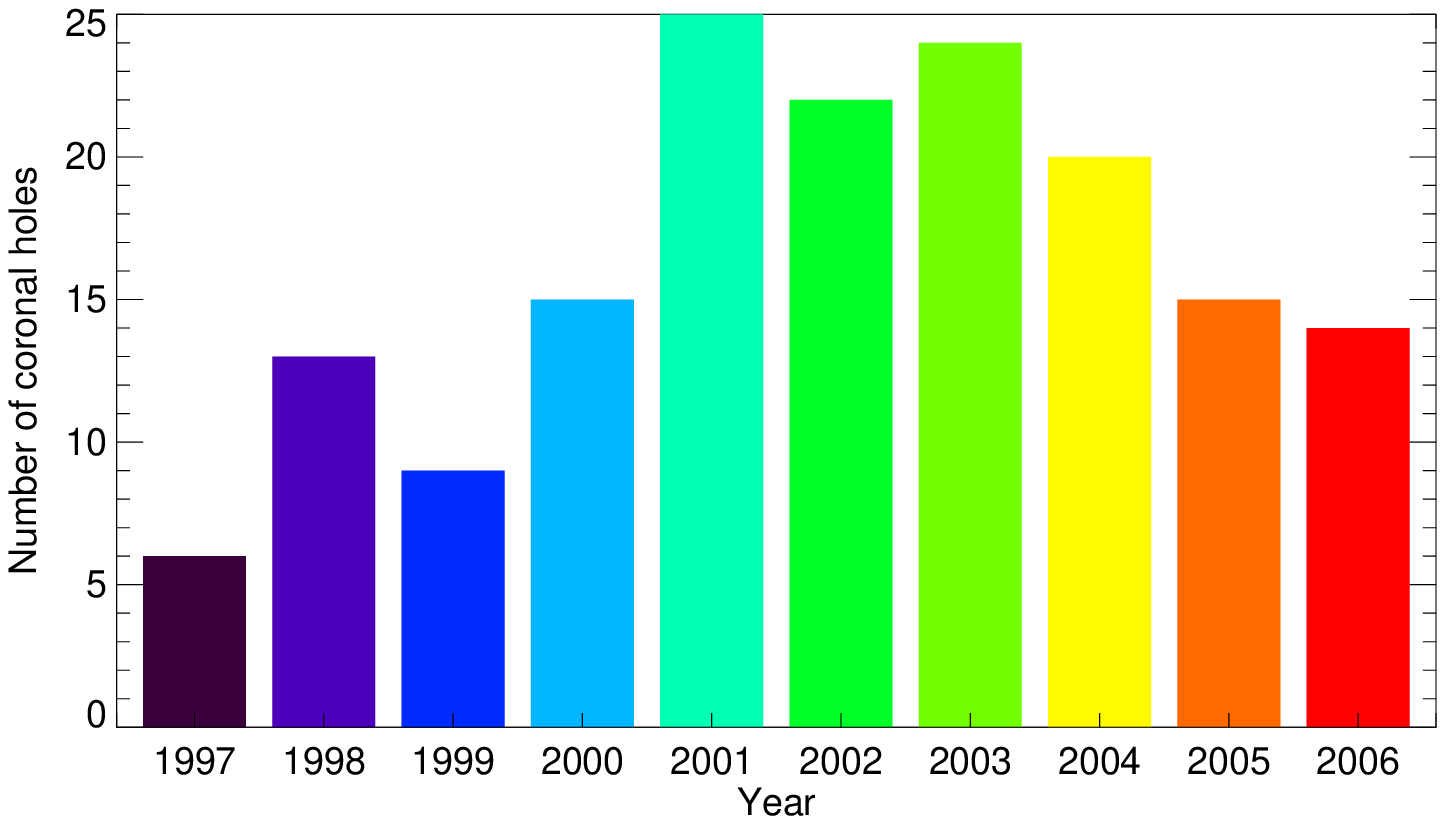} 
     \includegraphics[width=20.0pc,height=20.0pc]{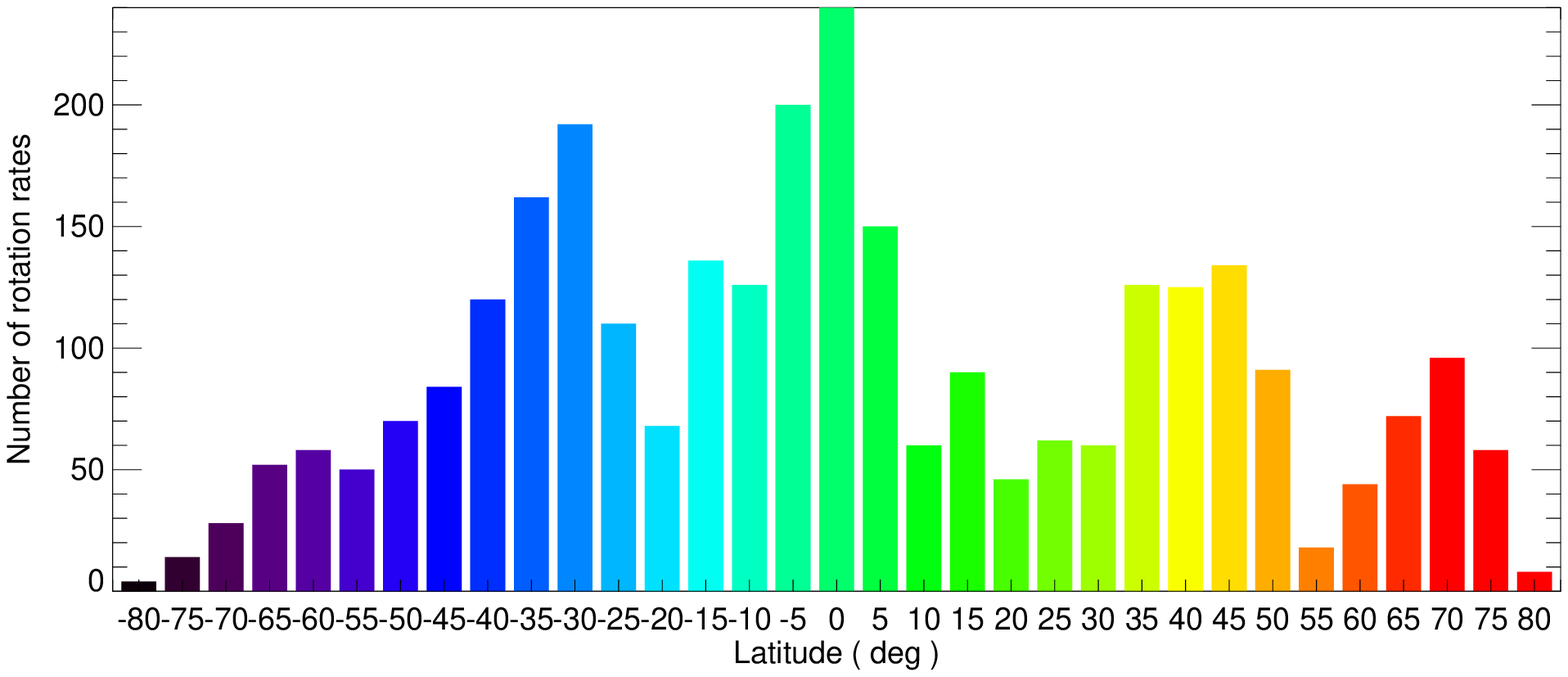}} 
\end{tabular}
    \caption{ For different years, Fig 2(a) illustrates occurrence
number of coronal holes. Whereas Fig 2(b) illustrates the number of rotation rates
used for estimation of average rotation rate of the coronal hole in a particular bin.
}
\end{center}
\end{figure}

\begin{figure}
\begin{center}
     Fig 3(a) \hskip 40ex  Fig 3(b)
\vskip -7ex
    \begin{tabular}{cc}
      {\includegraphics[width=20pc,height=20pc]{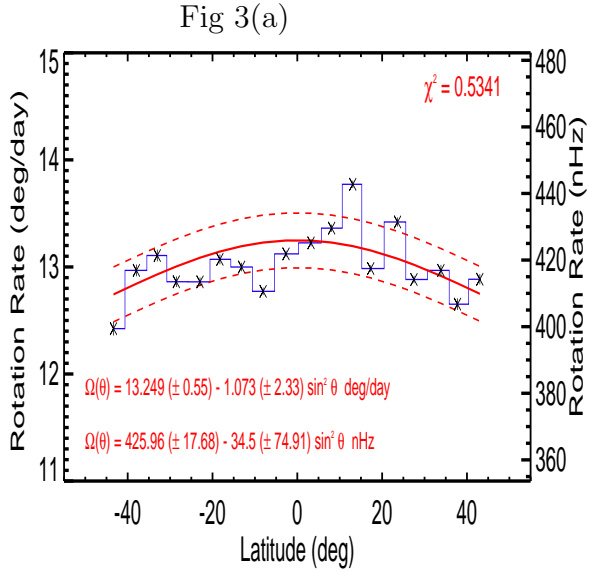}} &
      {\includegraphics[width=20pc,height=20pc]{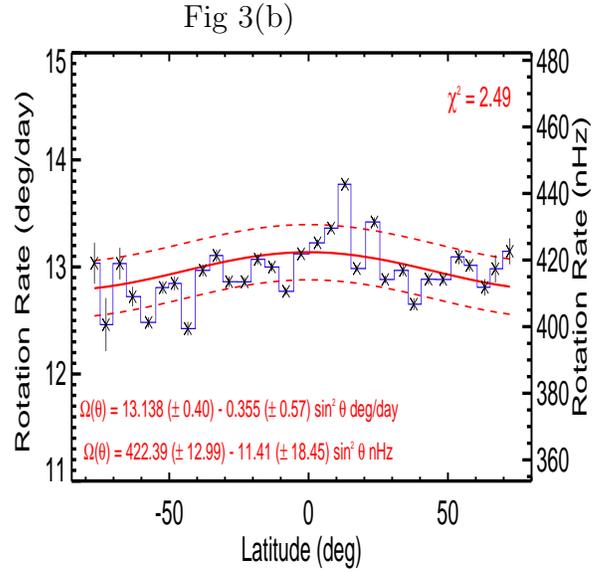}} \\
\end{tabular}
    \caption{Both the Figures illustrate the variation of rotation rates of
coronal holes for the longitude zones $65^{o}$ east to
$65^{o}$ west from the central meridian.
Fig 3(a) for the latitude zones between $45^{o}$ north to
$45^{o}$ south, whereas Fig 3(b) for the latitude zones from $80^{o}$ north to
$75^{o}$ south. In both the Figures blue bar curve
represents the observed rotation rates; red dashed lines represent
the one standard deviation (that is computed from all the data points)
error bands and, the red continuous line represents a least-square fit
of the observed data with a law of the form $\Omega(\theta) = \Omega_{0} + \Omega_{1} sin^2 (\theta)$.
The $\chi^{2}$ is a measure of goodness of fit over plotted on both
the plots.
}
\end{center}
\end{figure}

\begin{figure}
\begin{center}
     Fig 4(a) \hskip 40ex  Fig 4(b)
\vskip -7ex
    \begin{tabular}{cc}
      {\includegraphics[width=20pc,height=20pc]{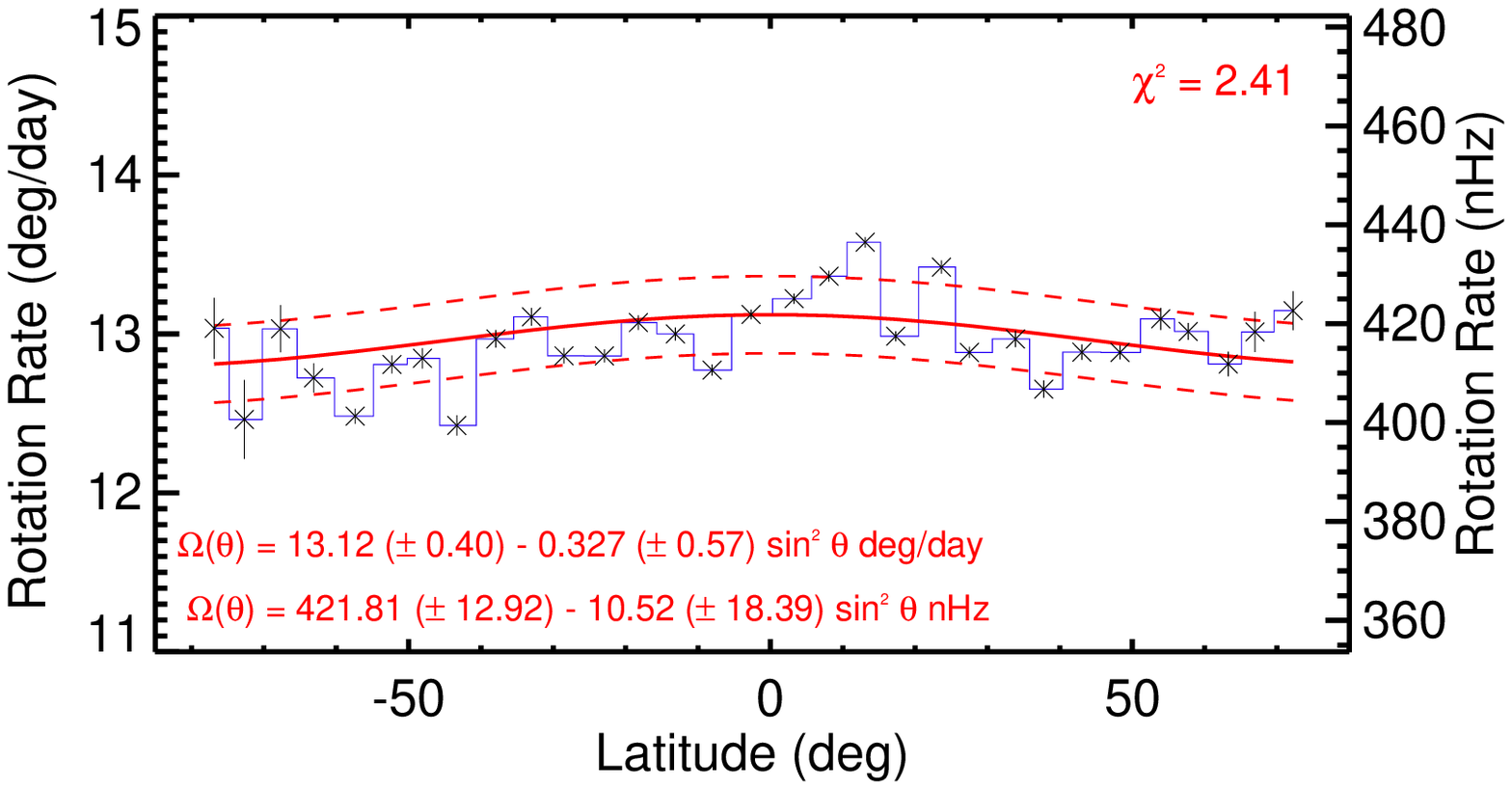}} &
      {\includegraphics[width=20pc,height=20pc]{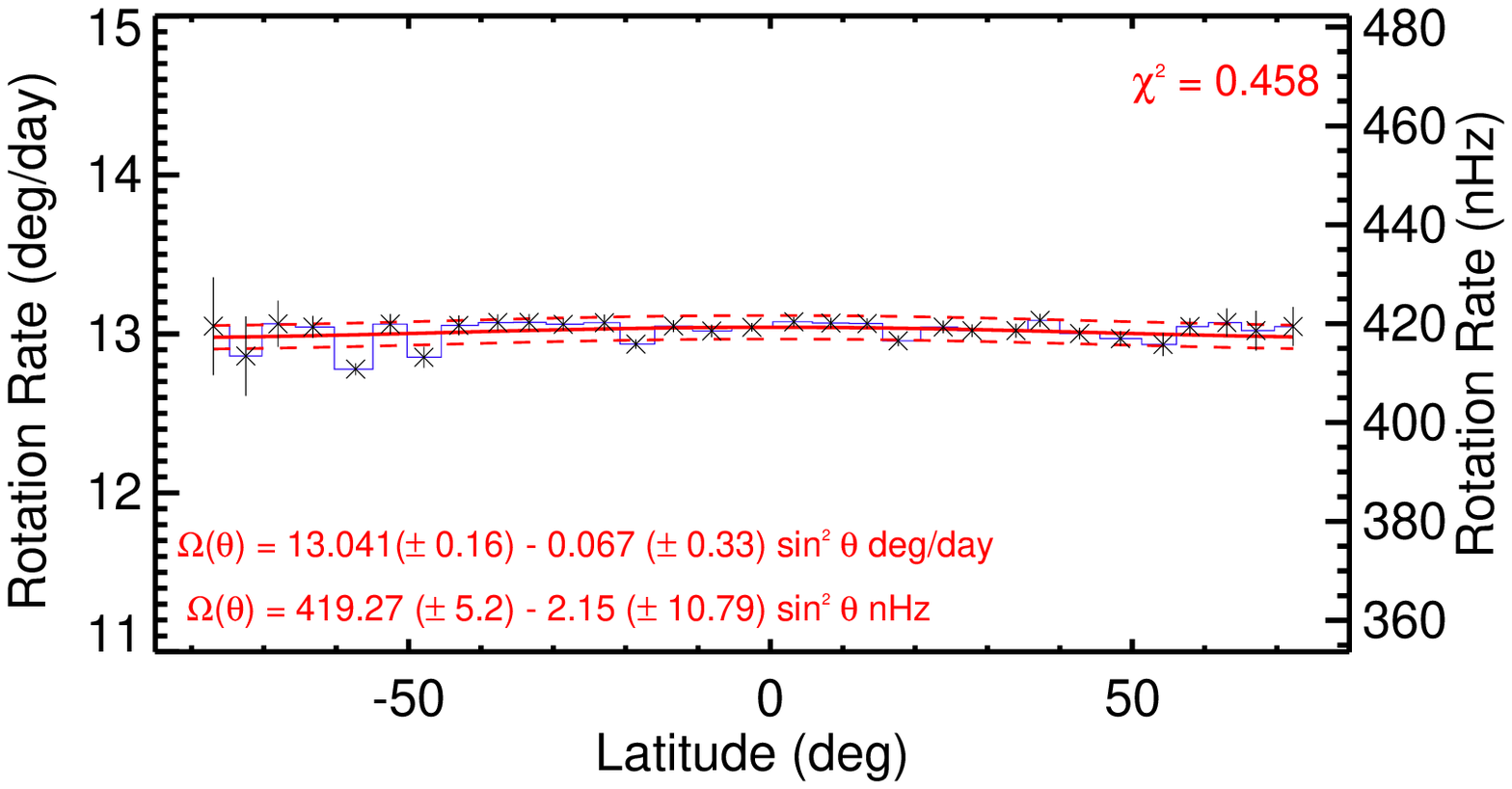}} \\
\end{tabular}
    \caption{ For the longitude zones $45^{o}$ east to
$45^{o}$ west from the central meridian, both the Figures
illustrate the variation
of rotation rate of coronal holes with respect to latitudes.
In case of Fig 4(b), projectional correction is applied.
In both the Figures blue bar curve
represents the observed rotation rates; red dashed lines represent
the one standard deviation (that is computed from all the data points)
error bands and, the red continuous line represents a least-square fit
of the observed data with a law of the form $\Omega(\theta) = \Omega_{0} + \Omega_{1} sin^2 (\theta)$.
The $\chi^{2}$ is a measure of goodness of fit over plotted on both
the plots.}
\end{center}
\end{figure}

\begin{figure}
\begin{center}
     Fig 5(a) \hskip 40ex  Fig 5(b)
\vskip -7ex
    \begin{tabular}{cc}
      {\includegraphics[width=20pc,height=20pc]{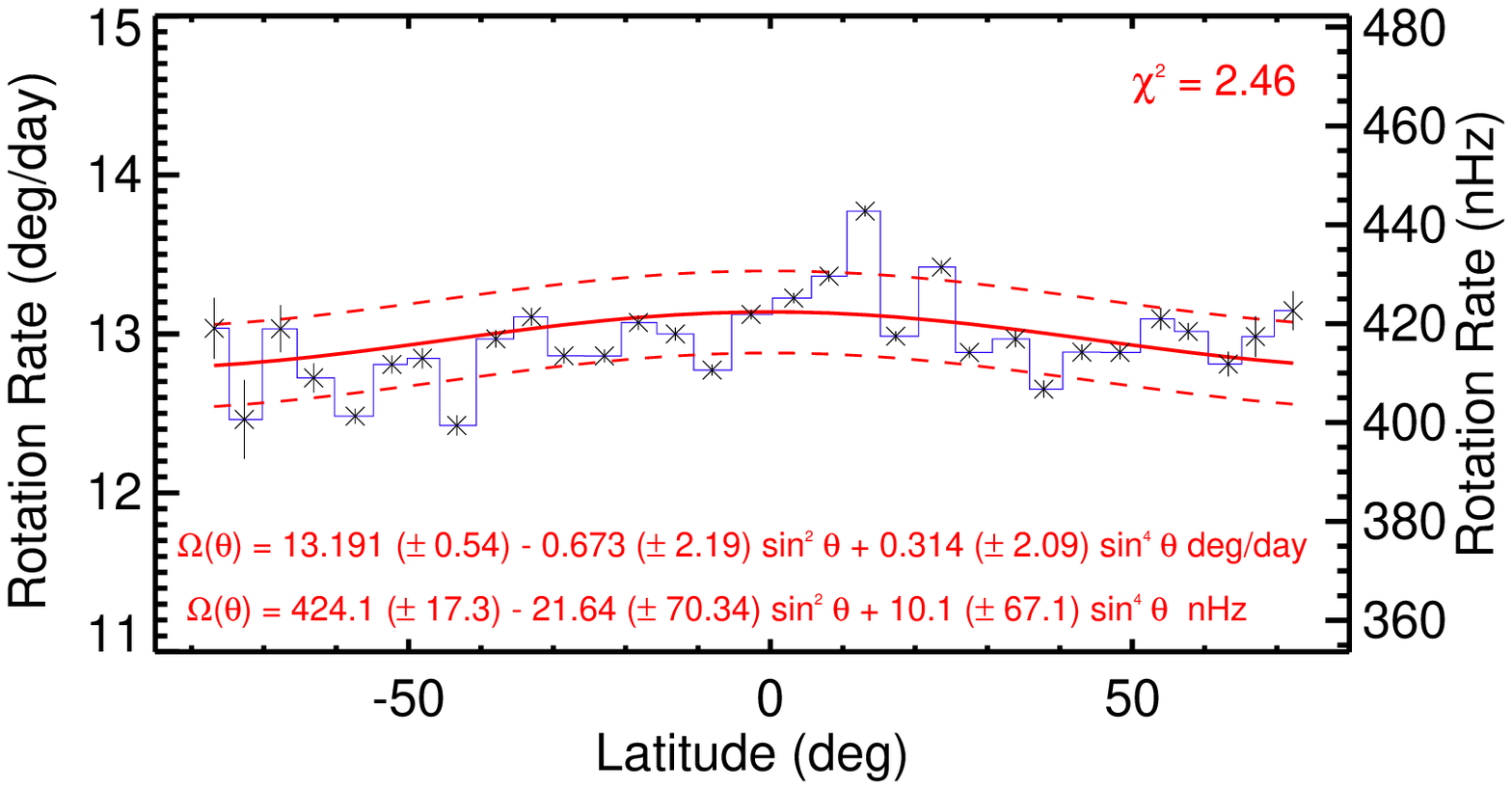}} &
      {\includegraphics[width=20pc,height=20pc]{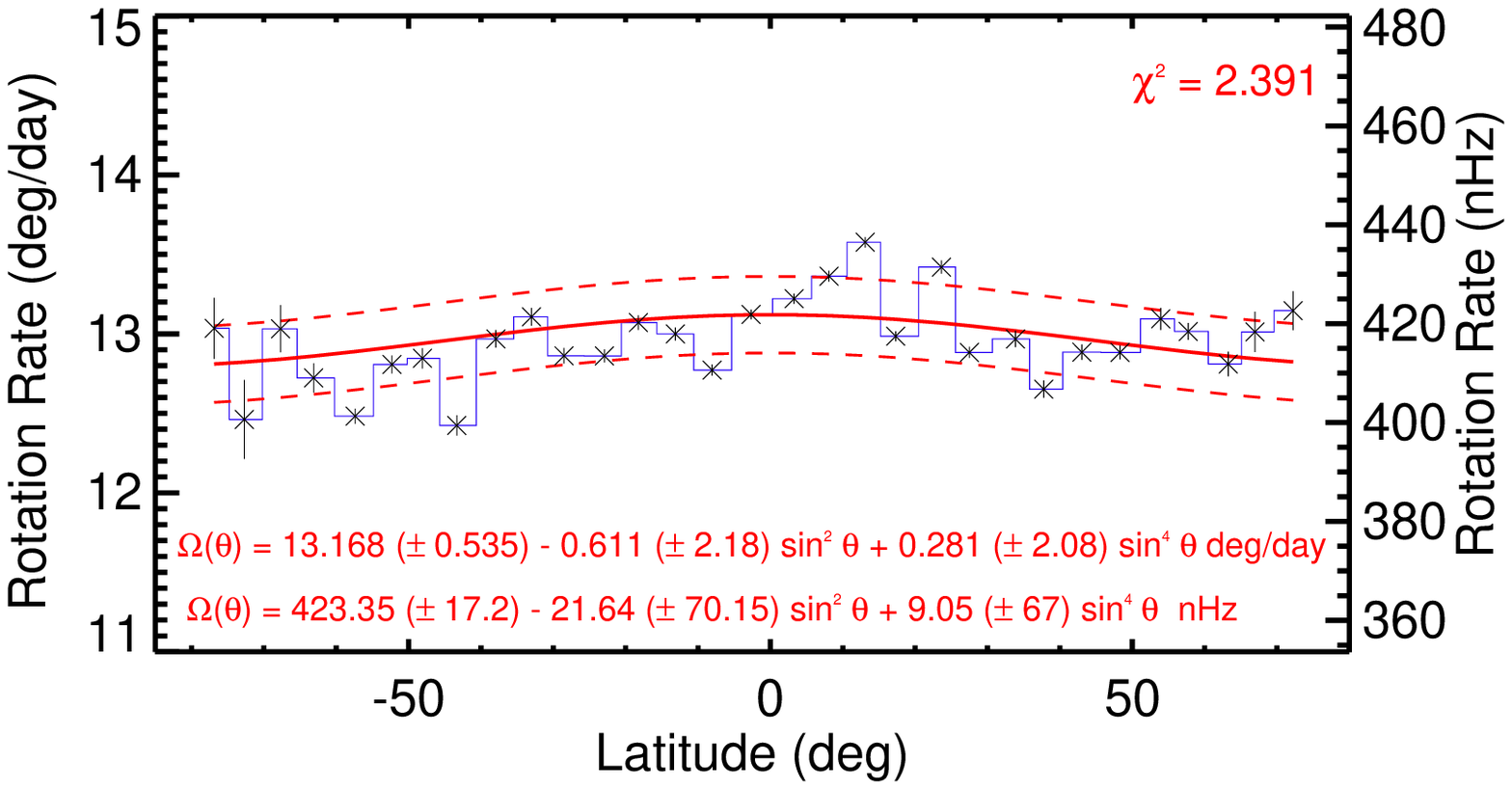}} \\
\end{tabular}
    \caption{ For the longitude zones $65^{o}$ east to
$65^{o}$ west from the central meridian, Fig 5(a) illustrates
the  variation of rotation rate of coronal holes with respect to latitudes.
Except constraint of $45^{o}$ east to
$45^{o}$ west from the central meridian, whereas Fig 5(b) represents
variation of rotation rate of coronal holes.
In both the Figures blue bar curve
represents the observed rotation rates; red dashed lines represent
the one standard deviation (that is computed from all the data points)
error bands and, the red continuous line represents a least-square fit
of the observed data with a law of the form $\Omega(\theta) = \Omega_{0} + 
\Omega_{1} sin^2 (\theta) + \Omega_{2} sin^4 (\theta)$.
The $\chi^{2}$ is a measure of goodness of fit over plotted on both
the plots.}
\end{center}
\end{figure}

\begin{figure}
\begin{center}
\includegraphics[width=20pc,height=20pc]{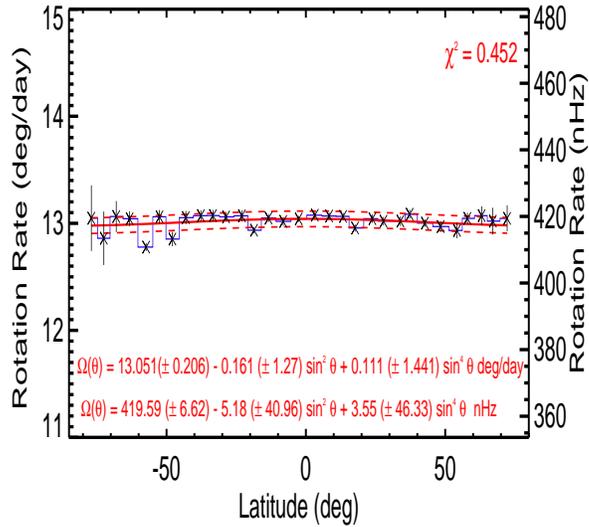}
    \caption{ For the same constraints of heliographic latitude and longitudes
as presented in Fig 5 and after taking into account the projectional effects,
this Figure illustrates the variation
of rotation rate of coronal holes with respect to latitudes.
Blue bar curve
represents the observed rotation rate; red dashed lines represent
the one standard deviation (that is computed from all the data points)
error bands and, the red continuous line represents a least-square fit of the form
$\Omega(\theta) = \Omega_{0} + \Omega_{1} sin^2 (\theta) +  \Omega_{2} sin^4 (\theta)$.
The $\chi^{2}$ is a measure of goodness of fit.}
\end{center}
\end{figure}

\begin{figure}
\begin{center}
     1997(a) \hskip 40ex  1997(b)
\vskip -6.5ex
    \begin{tabular}{cc}
      {\includegraphics[width=18pc,height=18pc]{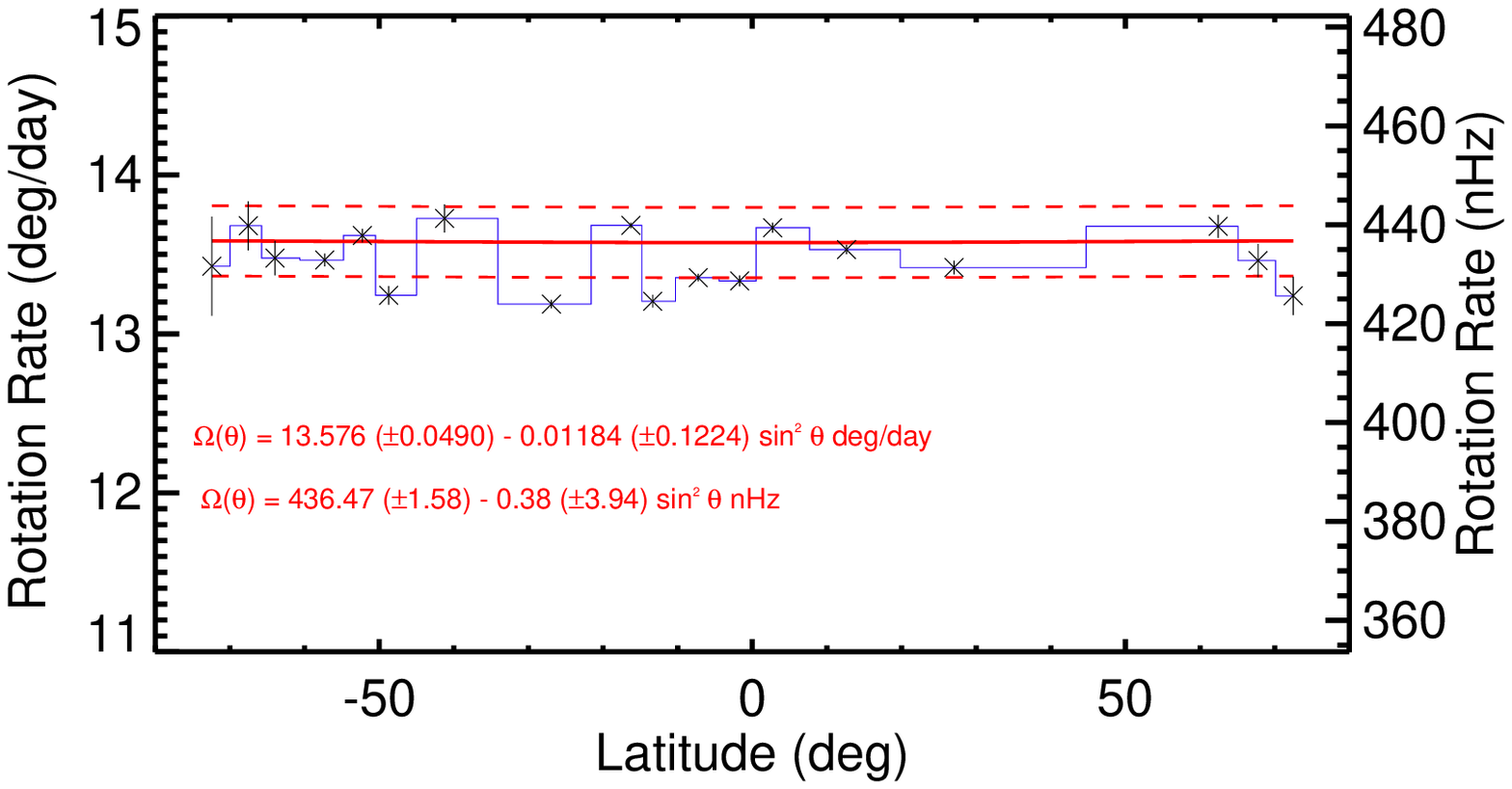}} &
      {\includegraphics[width=18pc,height=18pc]{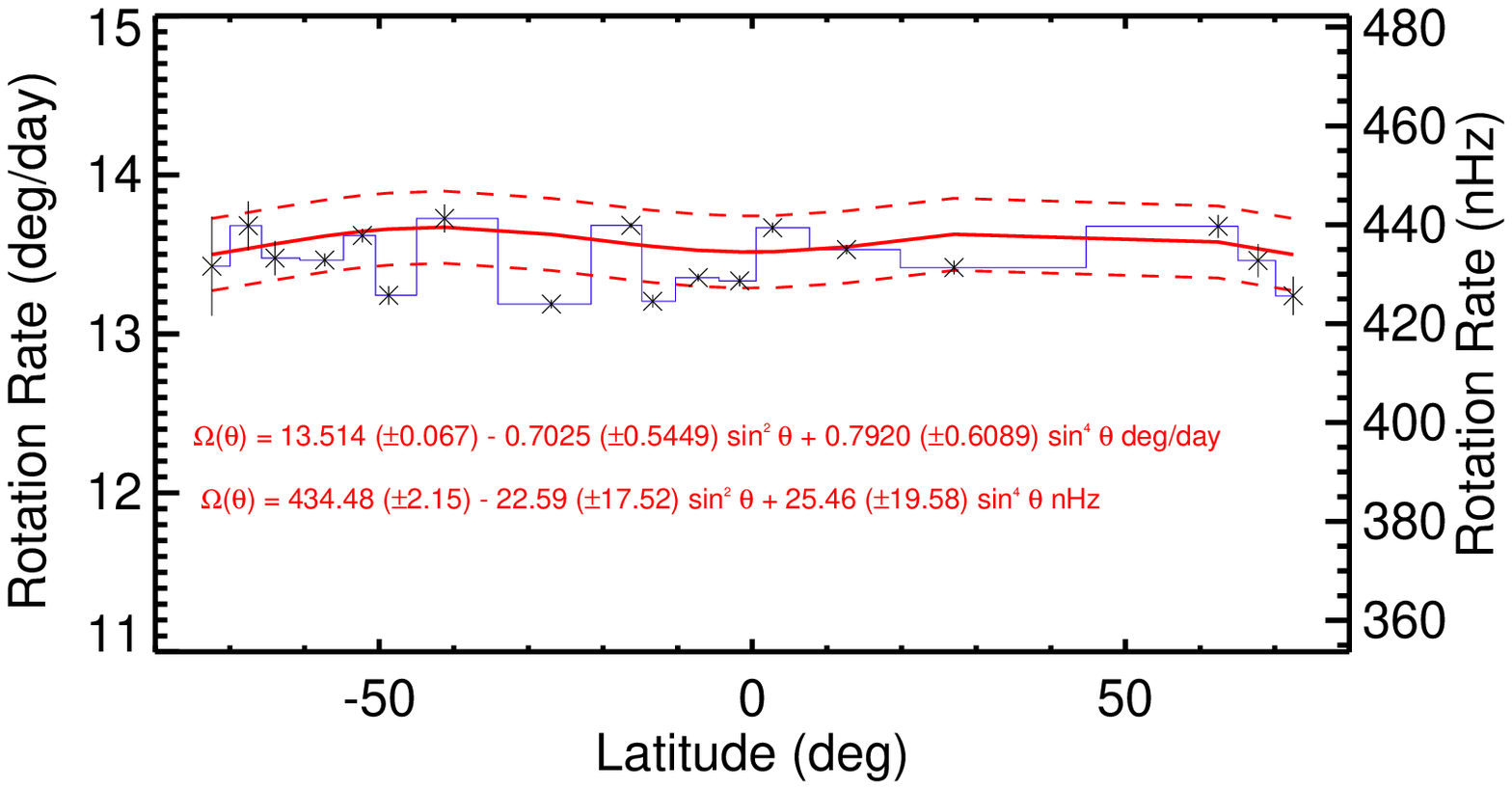}} \\
\end{tabular}
     1998(a) \hskip 40ex  1998(b)
\vskip -6.5ex
\begin{tabular}{cc}
      {\includegraphics[width=18pc,height=18pc]{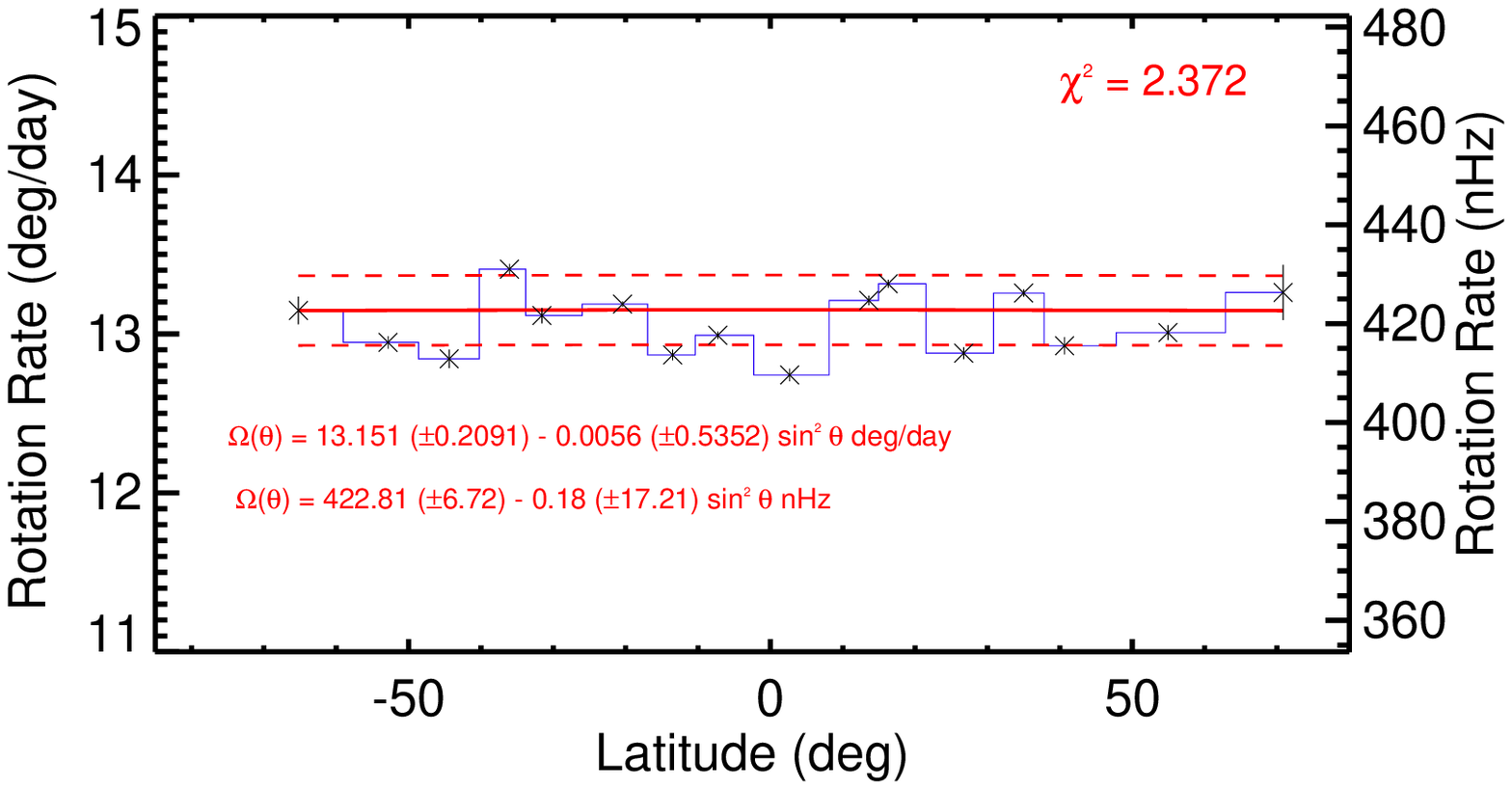}} &
     {\includegraphics[width=18pc,height=18pc]{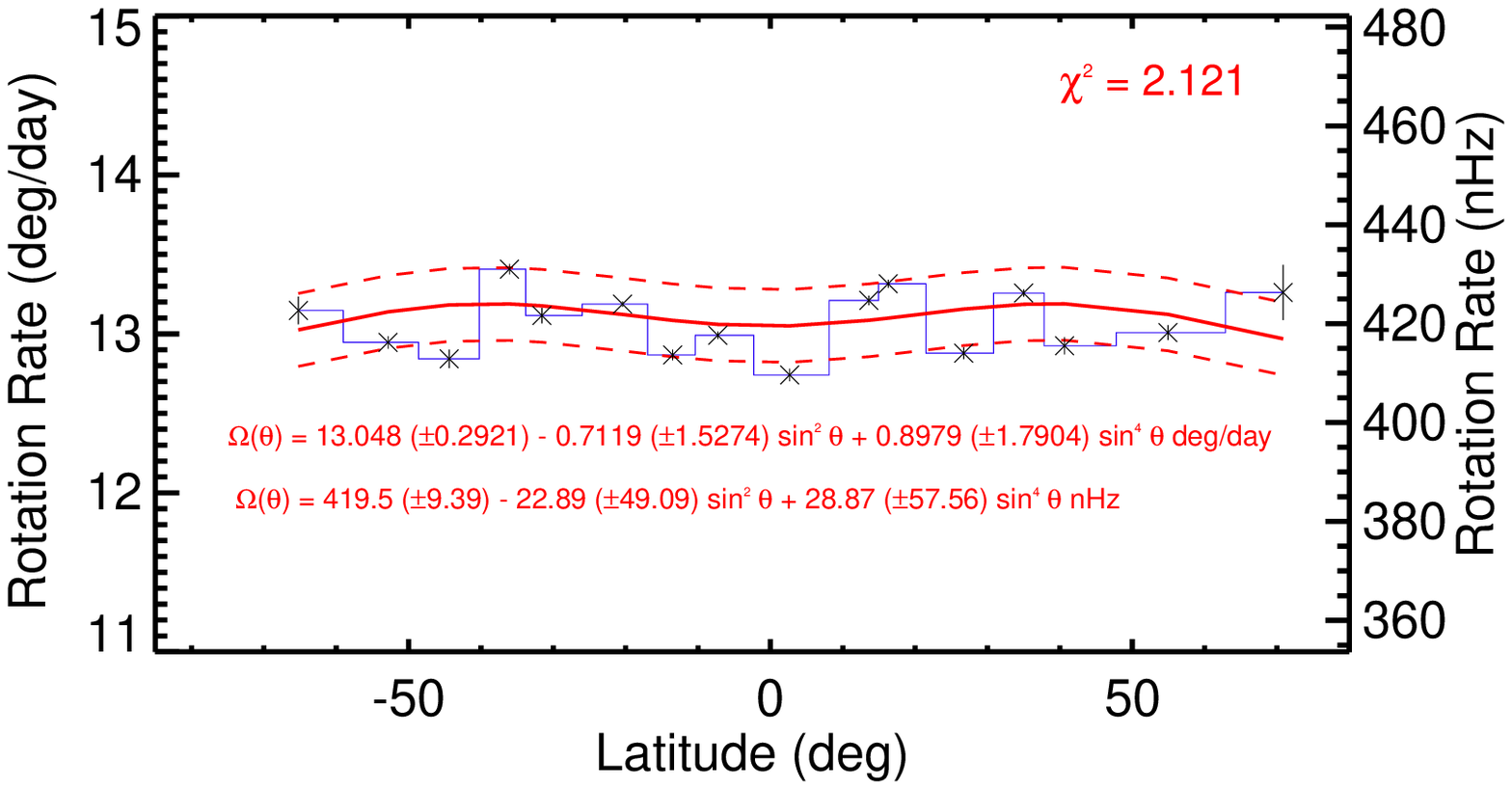}} \\
\end{tabular}
1999(a) \hskip 40ex  1999(b)
\vskip -6.5ex
\begin{tabular}{cc}
    {\includegraphics[width=18pc,height=18pc]{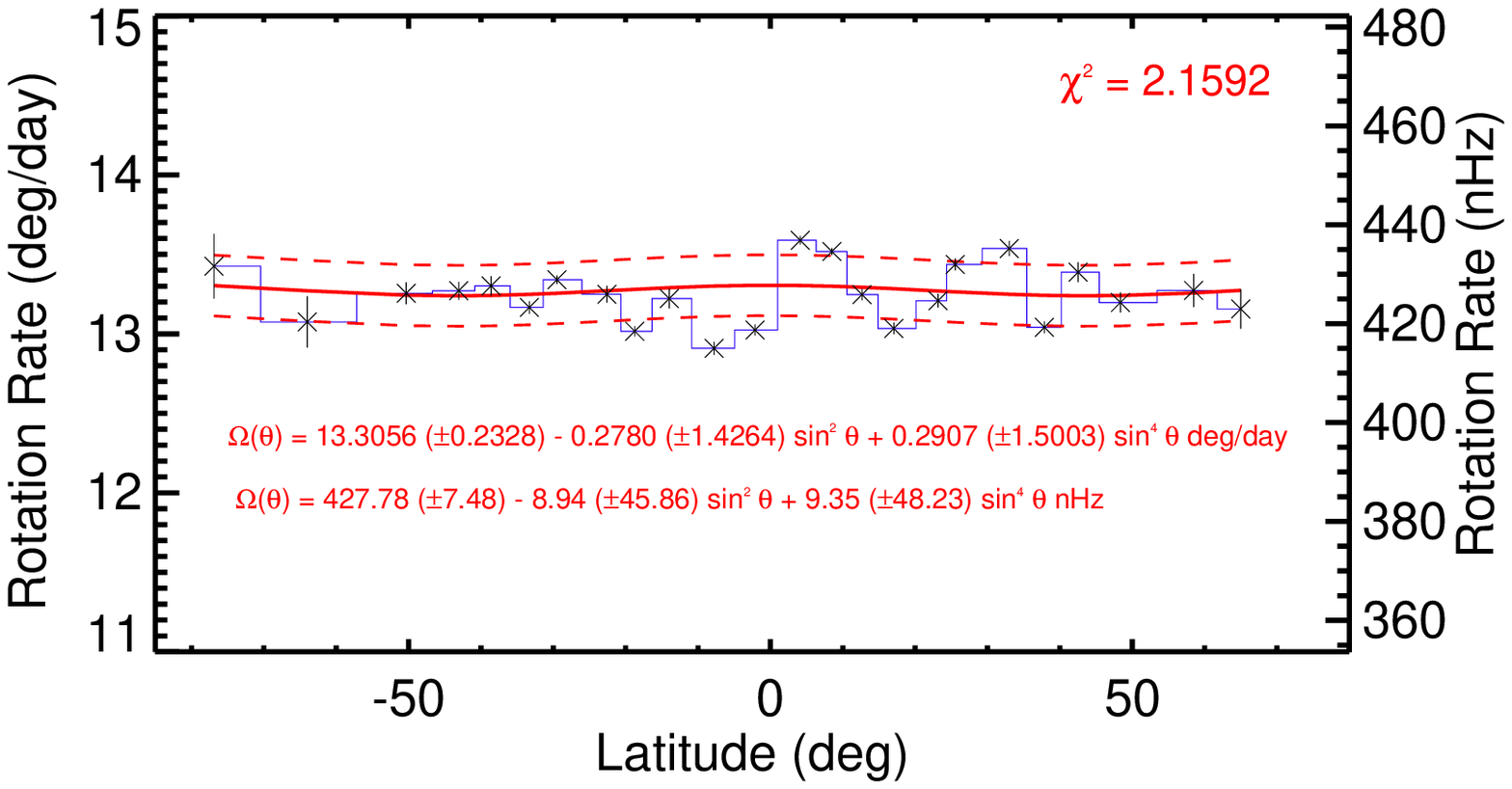}} &
      {\includegraphics[width=18pc,height=18pc]{overallfromn80_to_p75__1999.eps}} \\
\end{tabular}
    \caption{For the years 1997-1999, latitudinal variation of rotation rate of the coronal holes.
Left panel is fitted with a law up to $sin^{2} \theta$, whereas right
panel is fitted with a law up to $sin^{4} \theta$. }
\end{center}
\end{figure}

\begin{figure}
\begin{center}
    2000(a) \hskip 40ex  2000(b)
\vskip -6.5ex
    \begin{tabular}{cc}
      {\includegraphics[width=18pc,height=18pc]{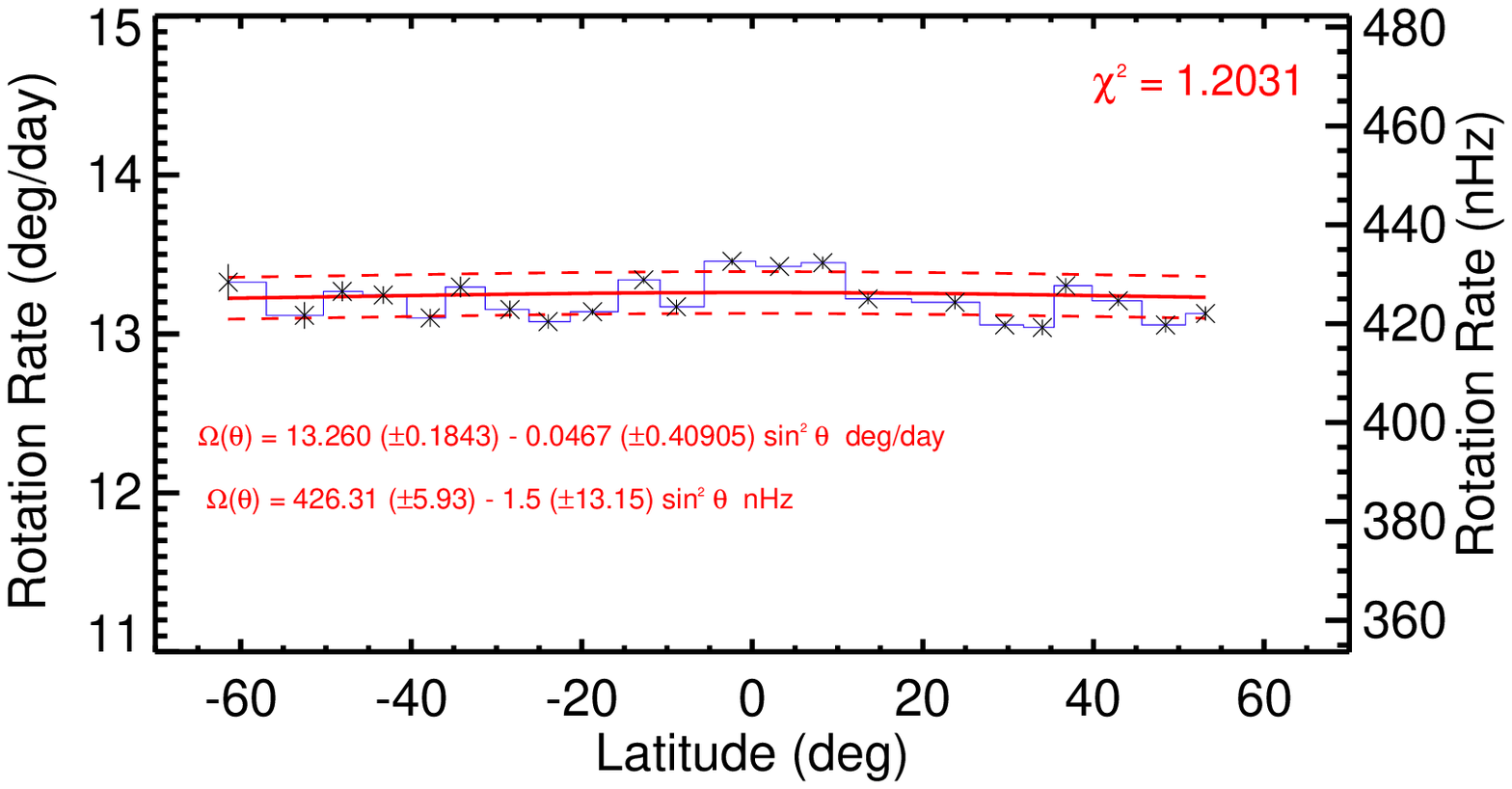}}&
      {\includegraphics[width=18pc,height=18pc]{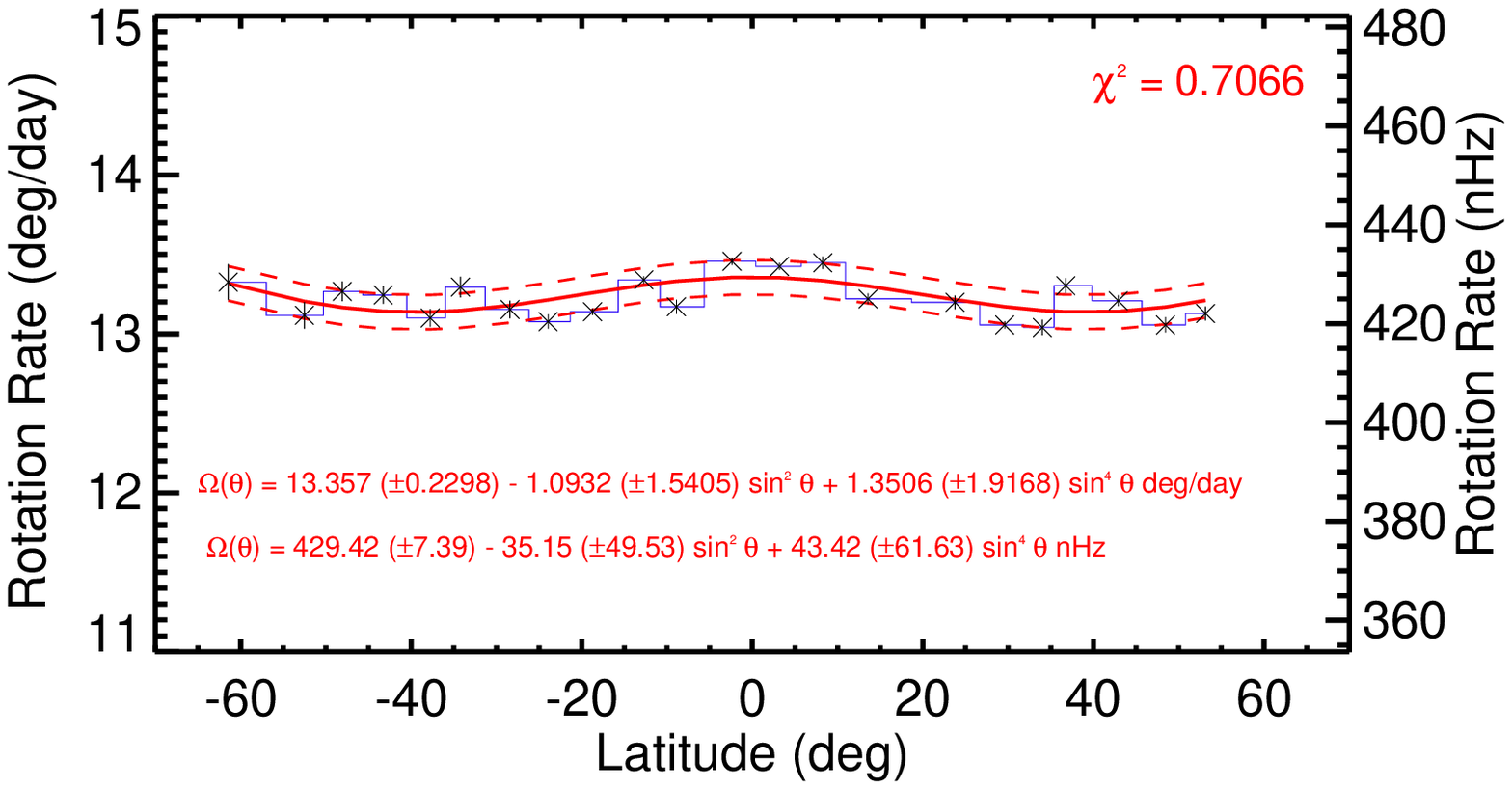}} \\
\end{tabular}
    2001(a) \hskip 40ex  2001(b)
\vskip -6.5ex
\begin{tabular}{cc}
      {\includegraphics[width=18pc,height=18pc]{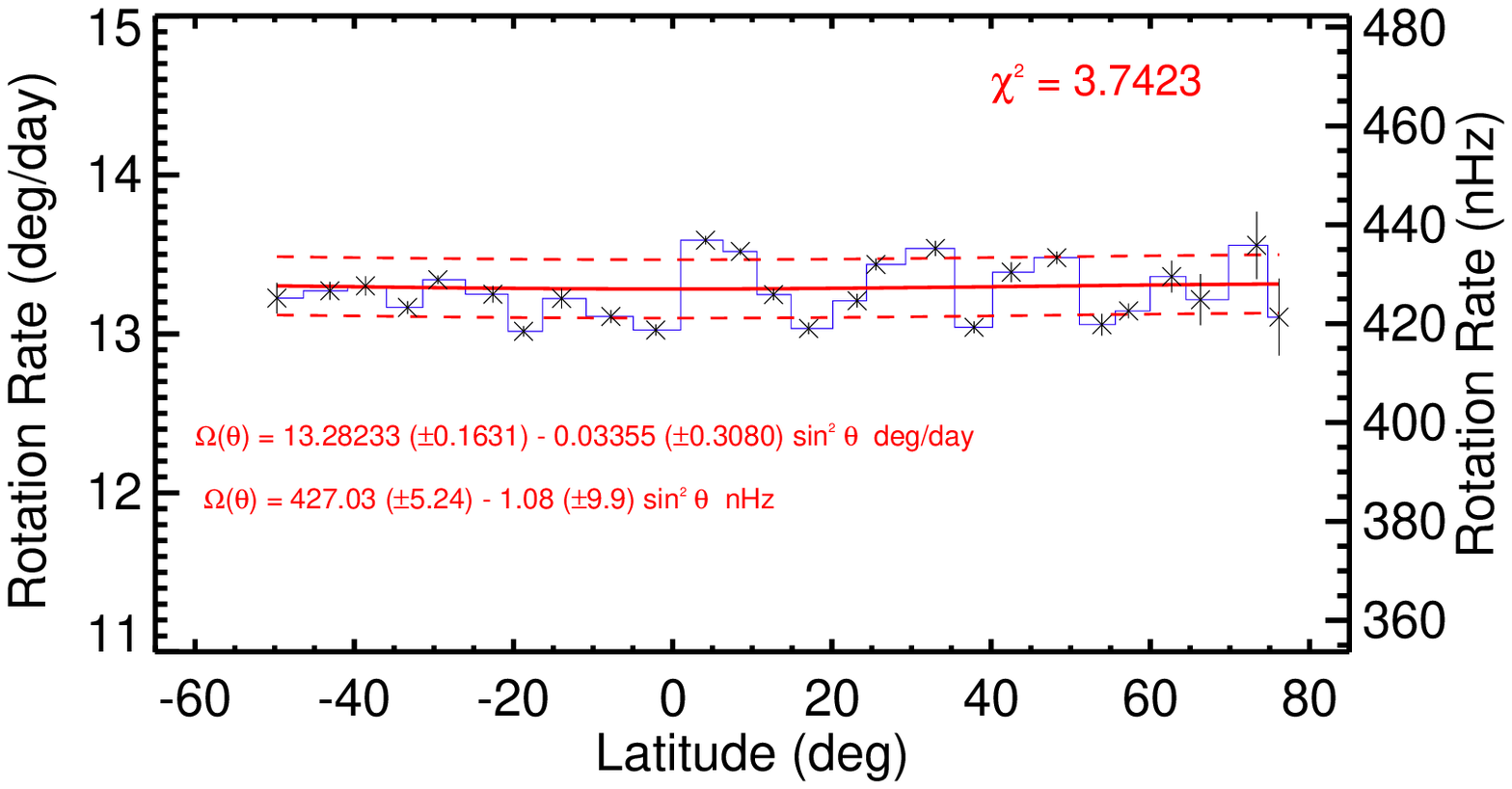}}&
      {\includegraphics[width=18pc,height=18pc]{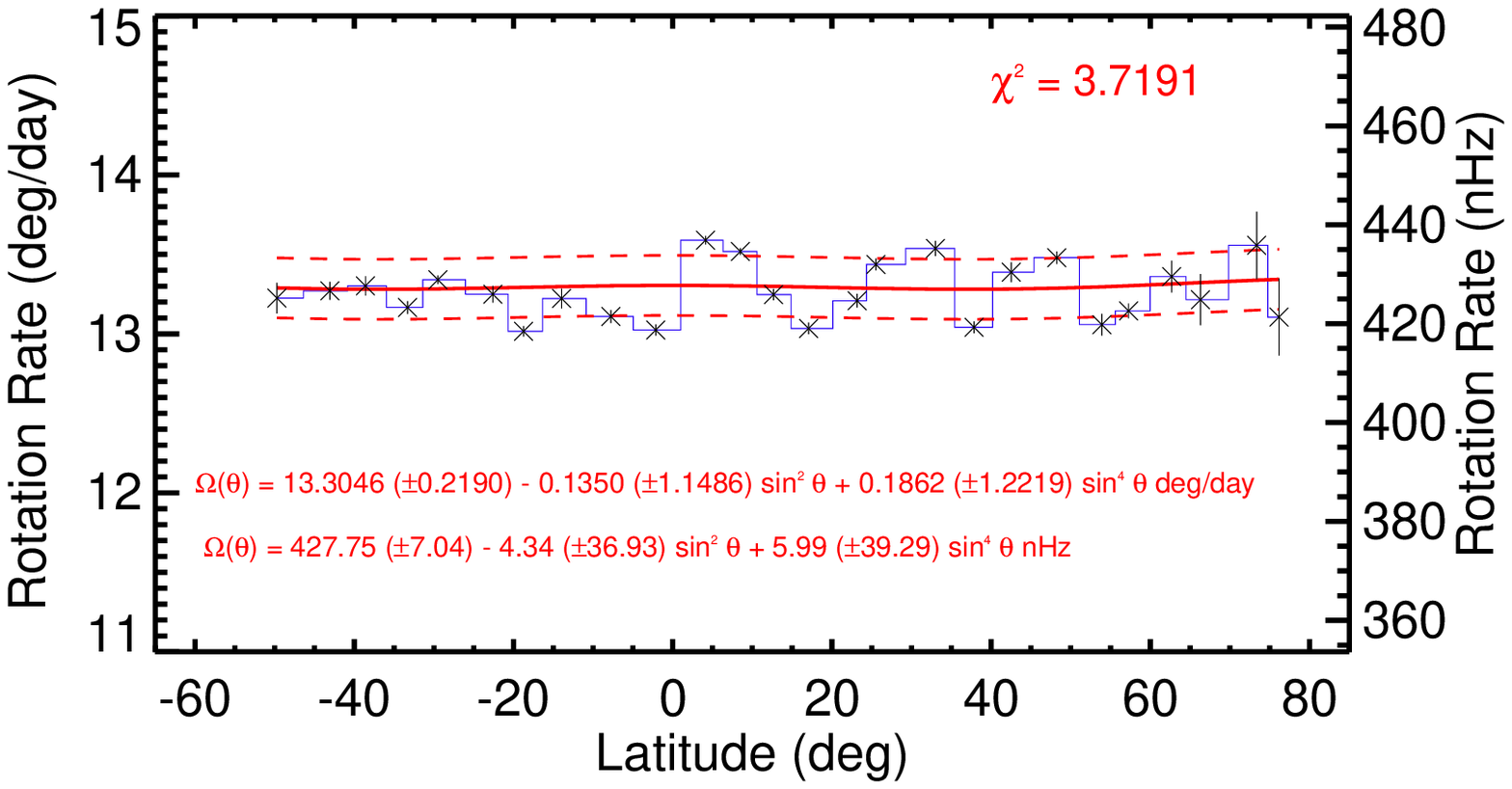}}\\
\end{tabular}
    2002(a) \hskip 40ex  2002(b)
\vskip -6.5ex
\begin{tabular}{cc}
      {\includegraphics[width=18pc,height=18pc]{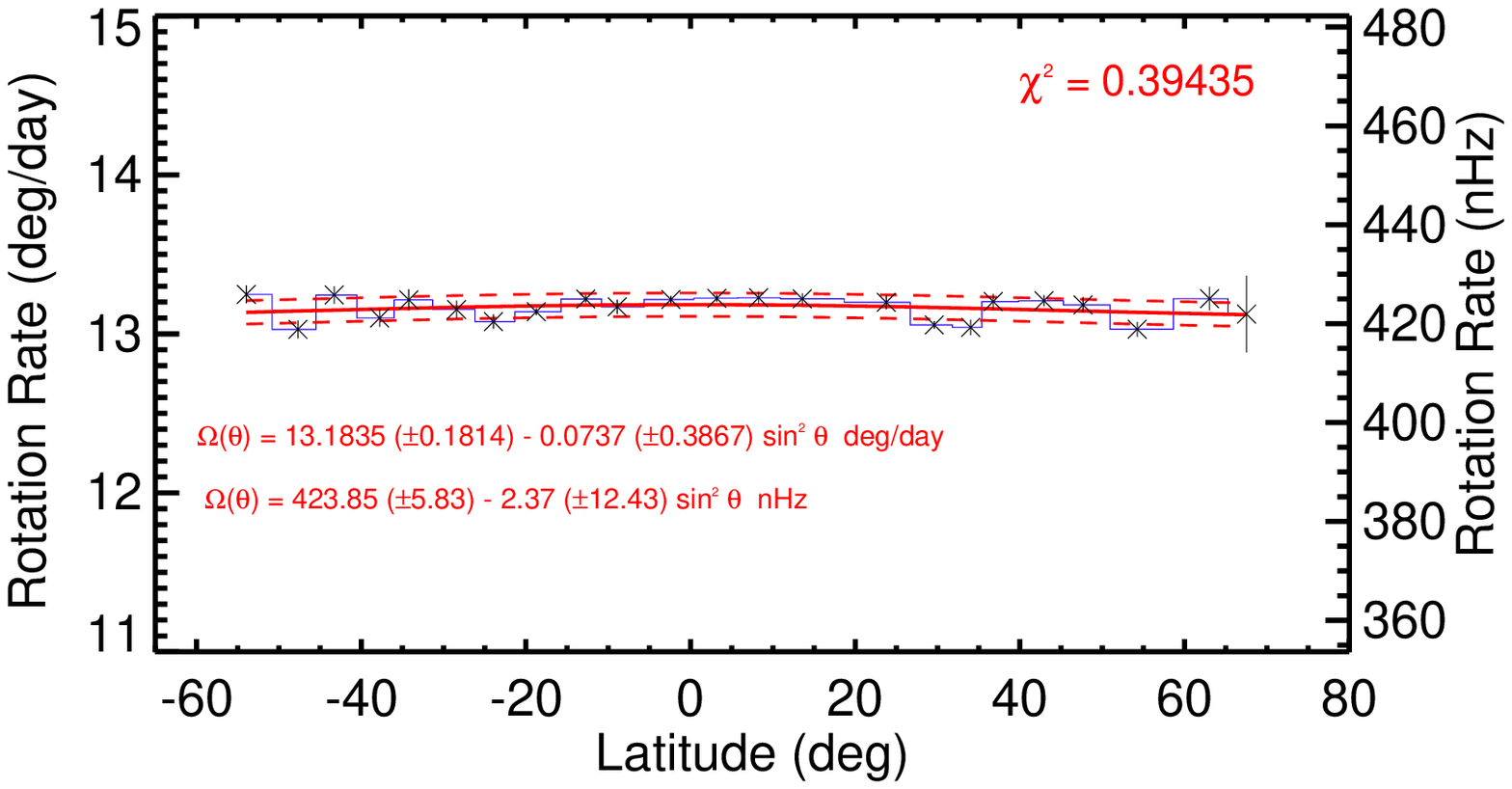}}&
      {\includegraphics[width=18pc,height=18pc]{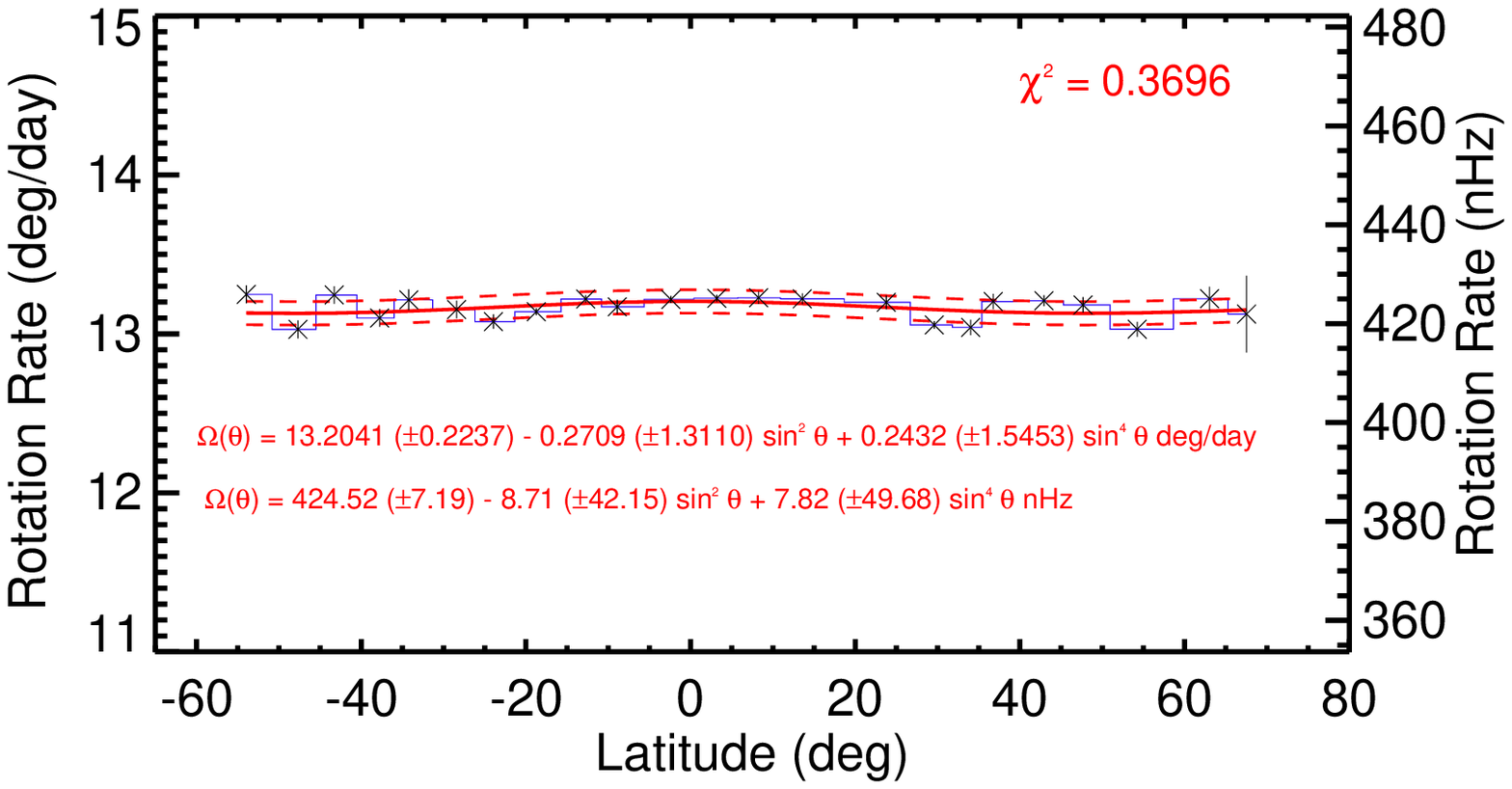}}\\
\end{tabular}
    \caption{For the years 2000-2002, latitudinal variation of rotation rate of the coronal holes.
Left panel is fitted with a law up to $sin^{2} \theta$, whereas right
panel is fitted with a law up to $sin^{4} \theta$}
 \end{center}
\end{figure}

\begin{figure}
\begin{center}
    2003(a) \hskip 40ex 2003(b)
\vskip -6.5ex
    \begin{tabular}{cc}
      {\includegraphics[width=18pc,height=18pc]{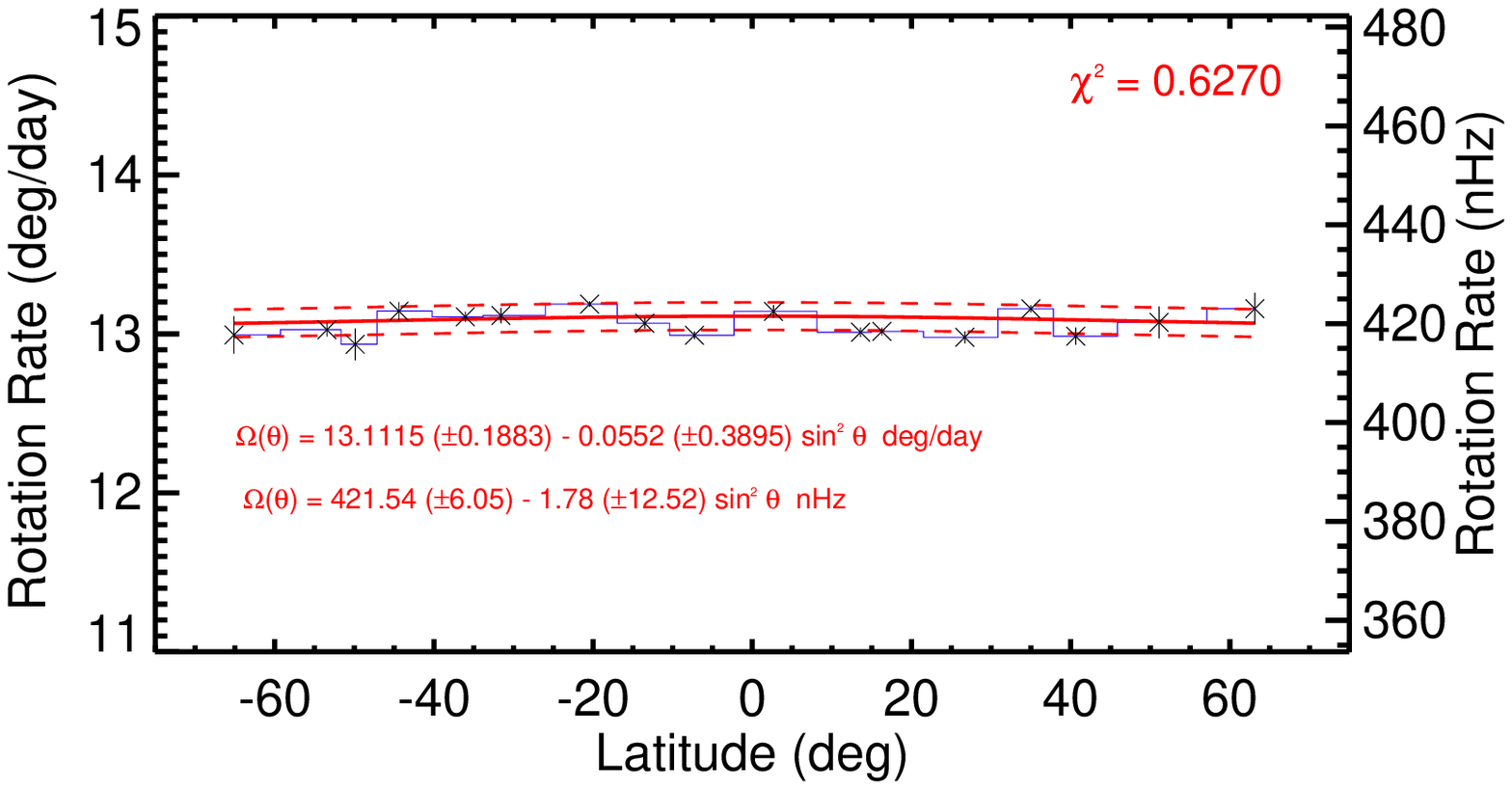}}&
      {\includegraphics[width=18pc,height=18pc]{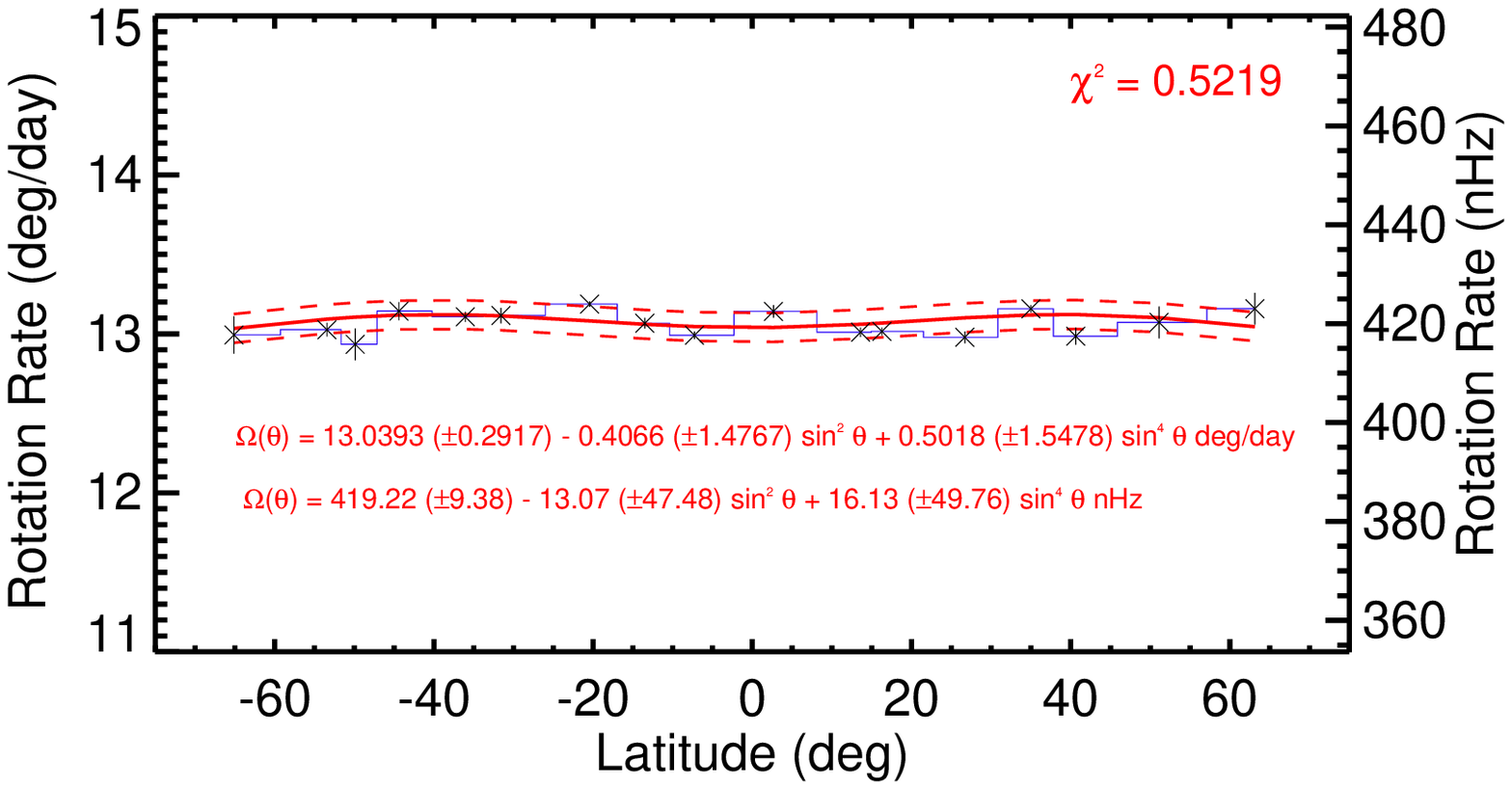}} \\
\end{tabular}
    2004(a) \hskip 40ex  2004(b)
\vskip -6.5ex
\begin{tabular}{cc}
      {\includegraphics[width=18pc,height=18pc]{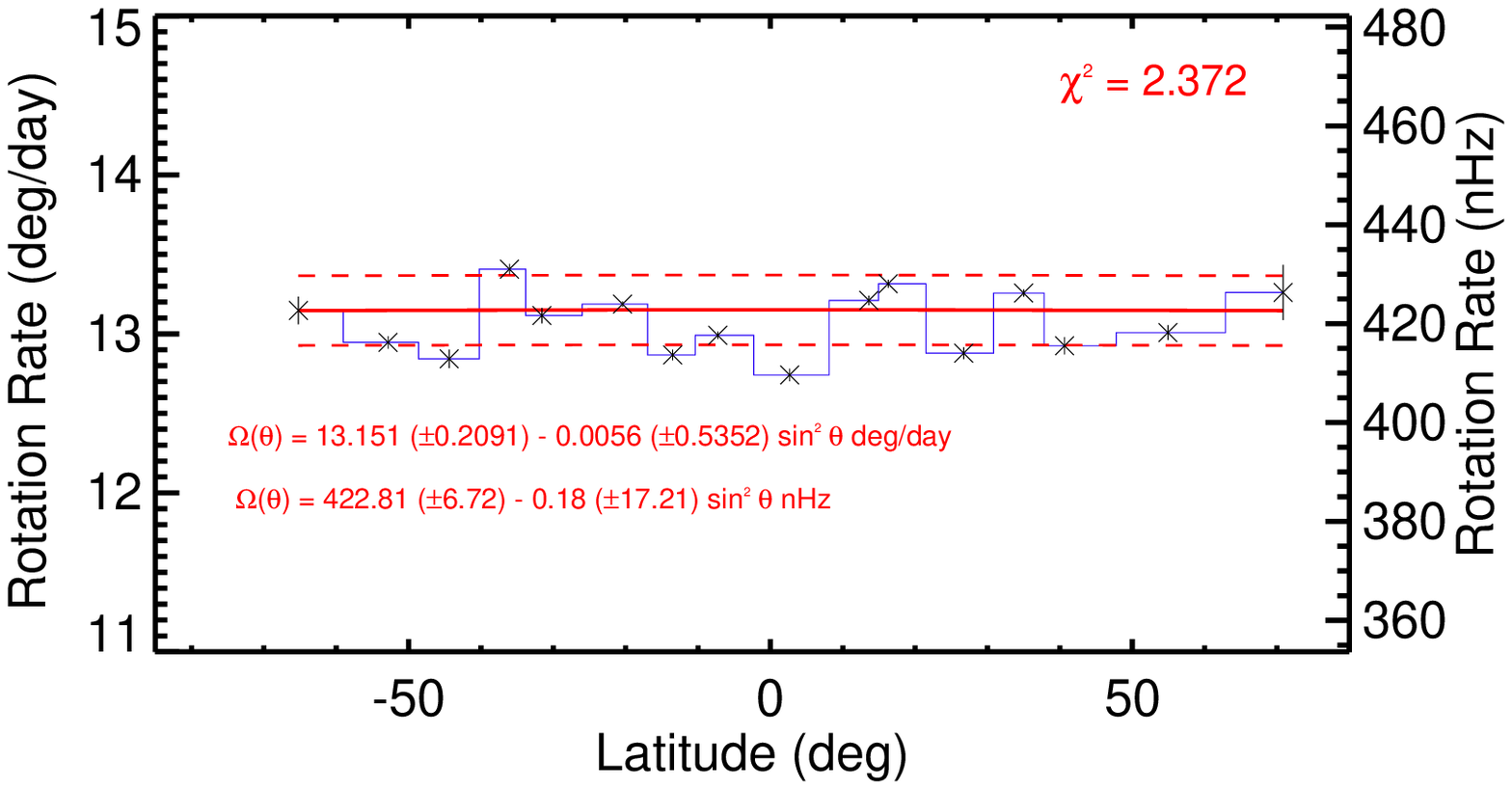}}&
      {\includegraphics[width=18pc,height=18pc]{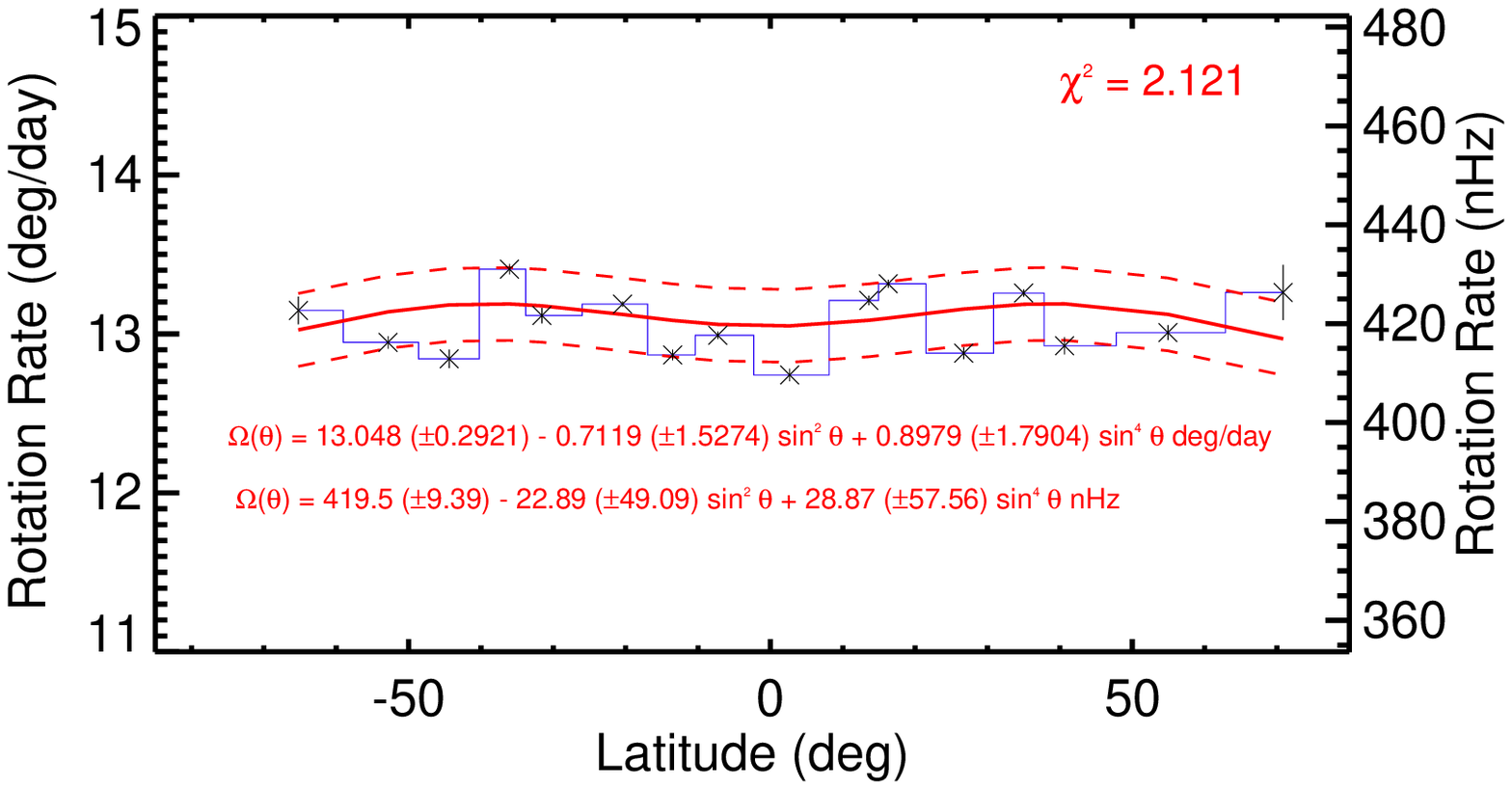}} \\
\end{tabular}
    2005(a) \hskip 40ex  2005(b)
\vskip -6.5ex
\begin{tabular}{cc}
      {\includegraphics[width=18pc,height=18pc]{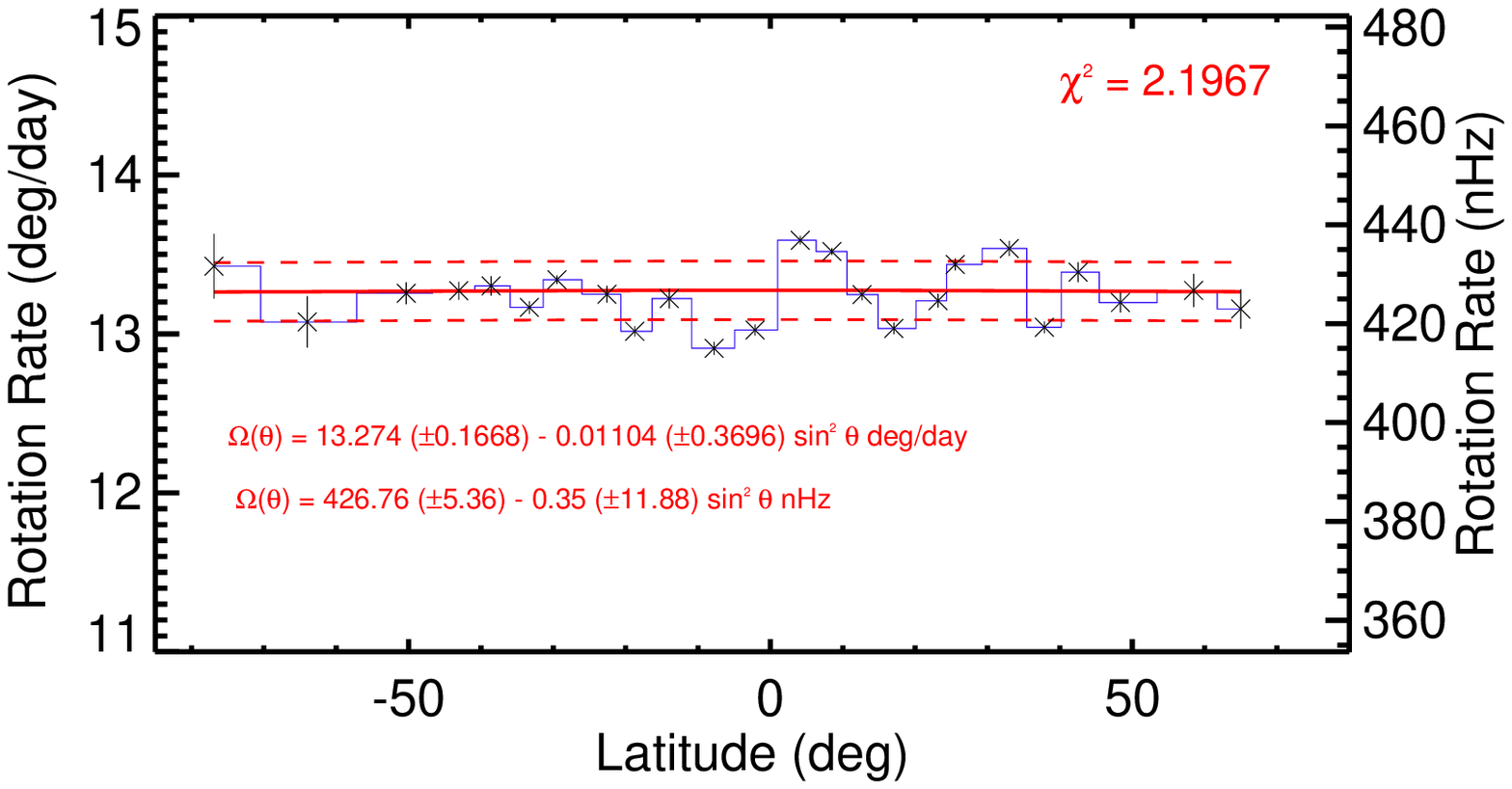}}&
      {\includegraphics[width=18pc,height=18pc]{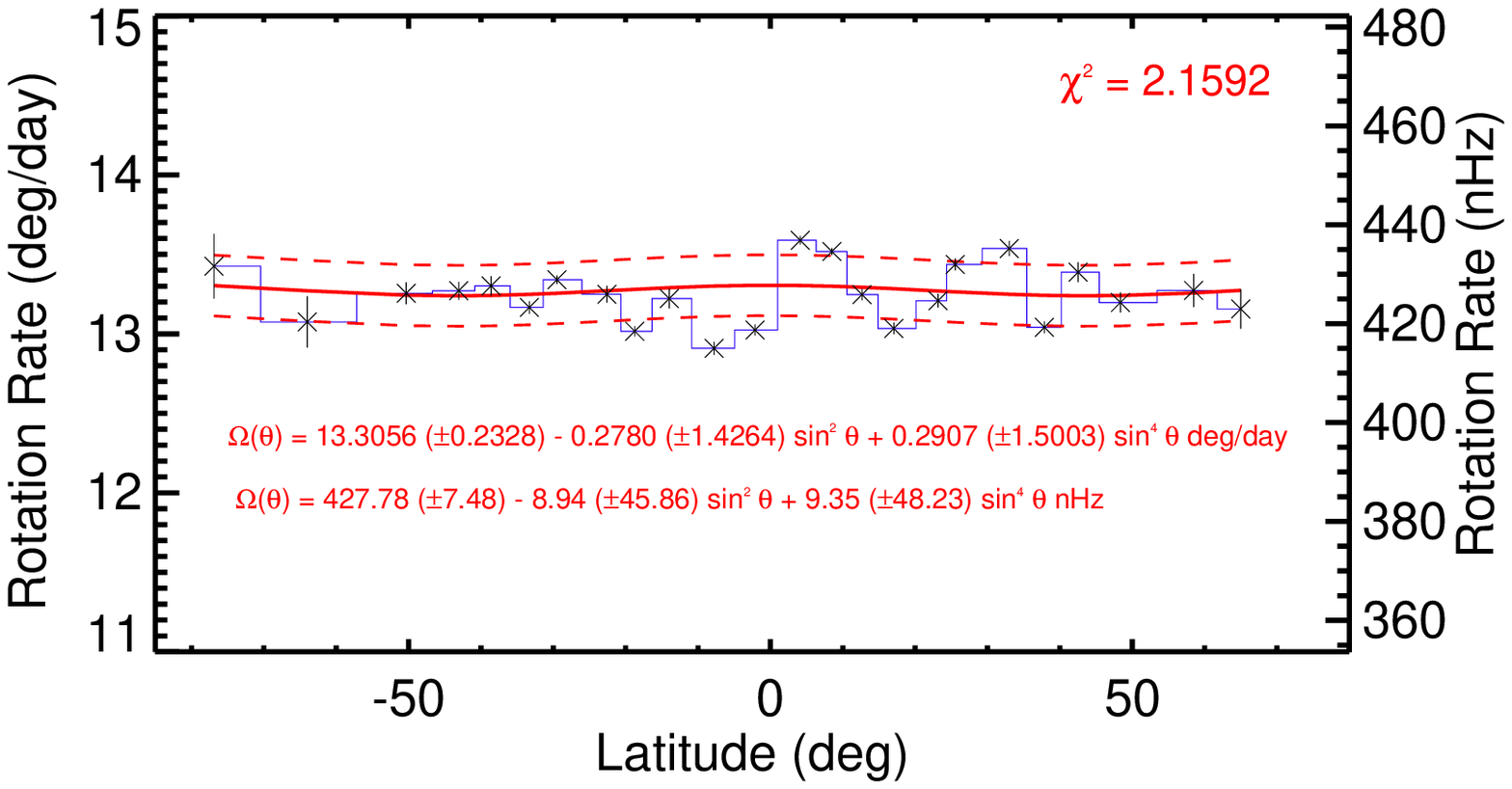}}\\
\end{tabular}
    \caption{For the years 2003-2005, latitudinal variation of rotation rate of the coronal holes.
Left panel is fitted with a law up to $sin^{2} \theta$, whereas right
panel is fitted with a law up to $sin^{4} \theta$.}
\end{center}
\end{figure}

\begin{figure}
\begin{center}
\vskip -6ex
    \begin{tabular}{cc}
      {\includegraphics[width=20pc,height=20pc]{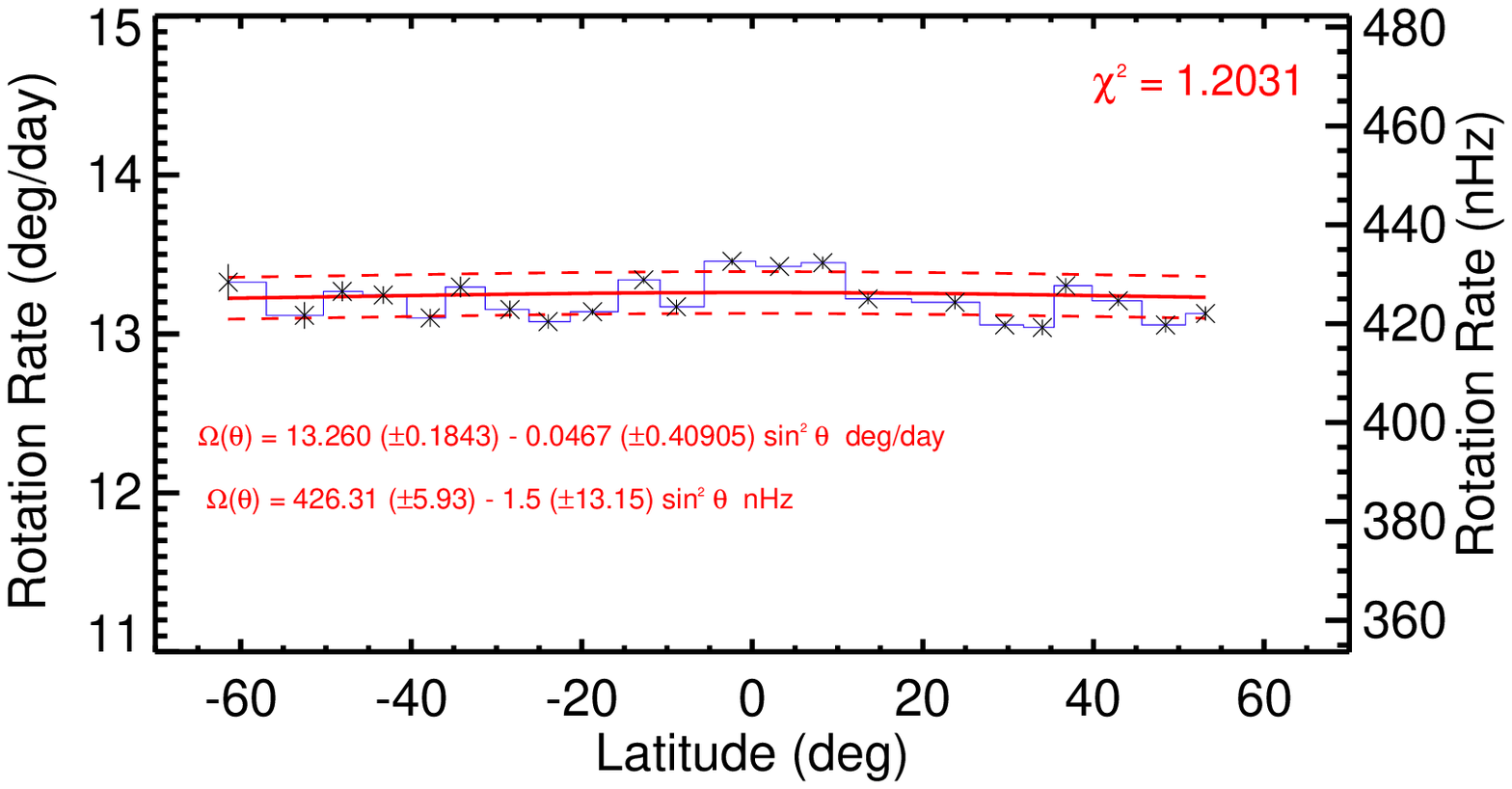}}&
      {\includegraphics[width=20pc,height=20pc]{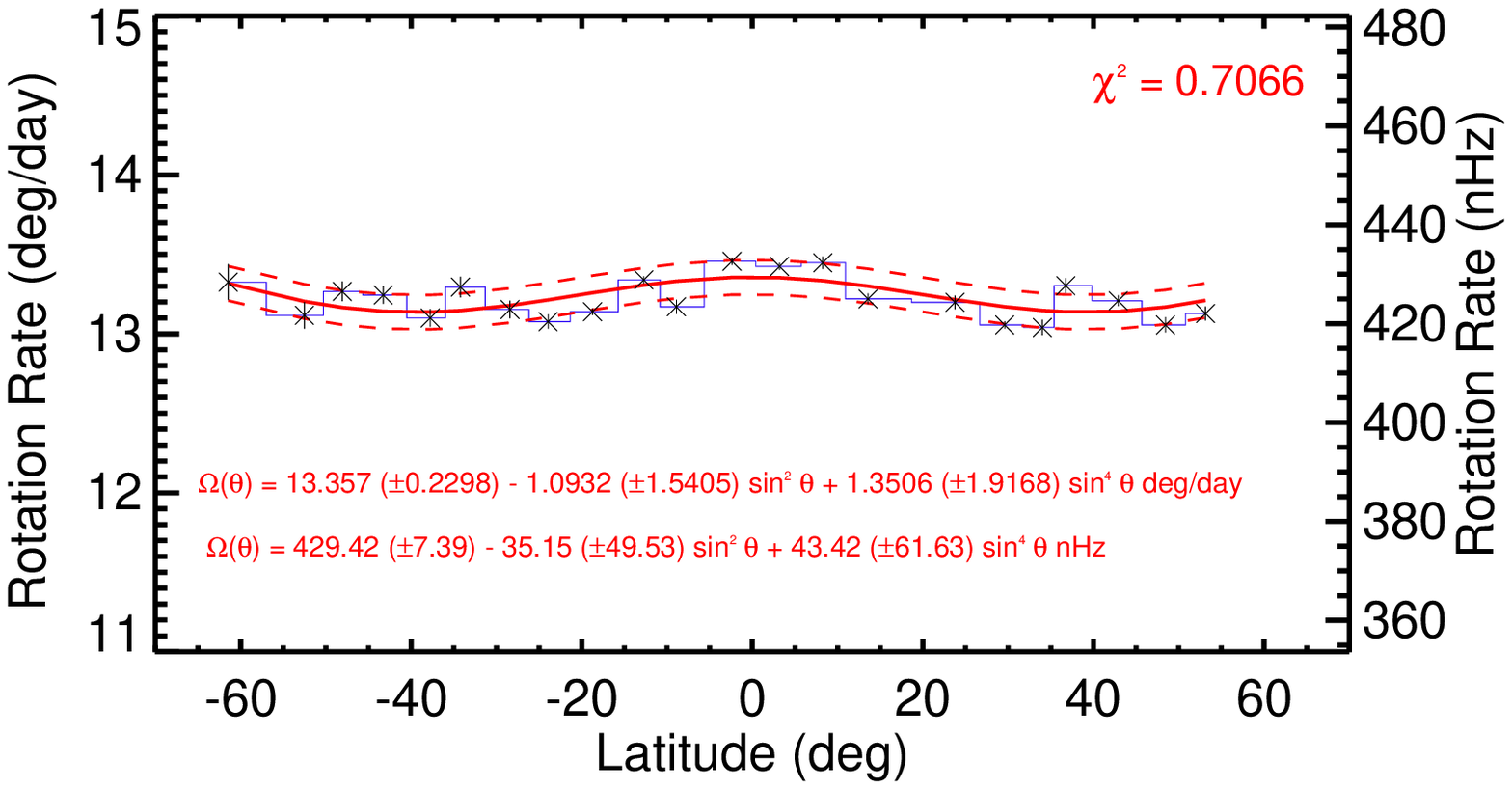}}\\
\end{tabular}
    \caption{ For the year 2006, latitudinal variation of rotation rate of the coronal holes.
Left panel is fitted with a law up to $sin^{2} \theta$, whereas right
panel is fitted with a law up to $sin^{4} \theta$.}
\end{center}
\end{figure}

\begin{figure}
\begin{center}
\vskip -6ex
    \begin{tabular}{cc}
      {\includegraphics[width=20pc,height=20pc]{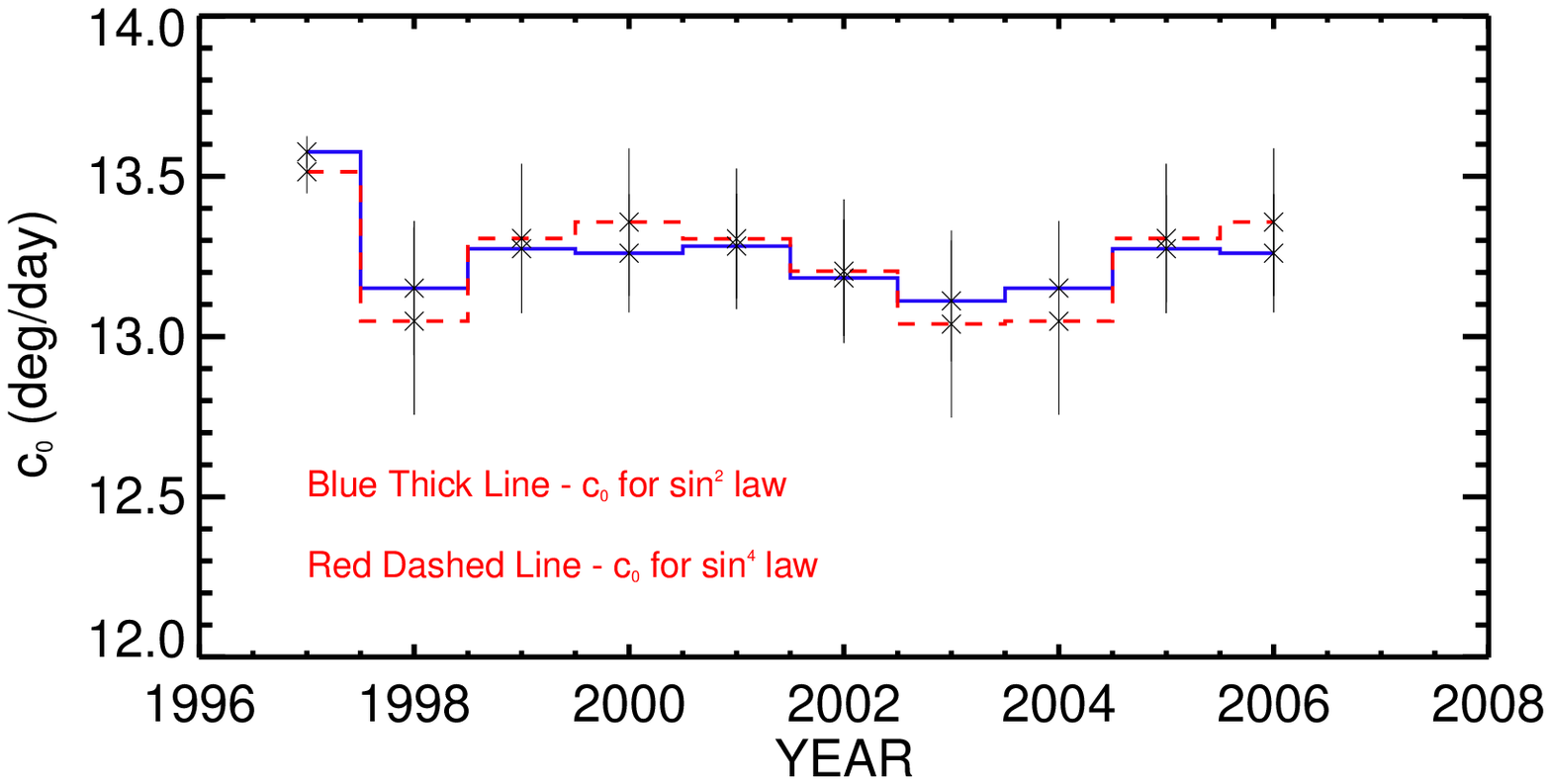}}&
      {\includegraphics[width=20pc,height=20pc]{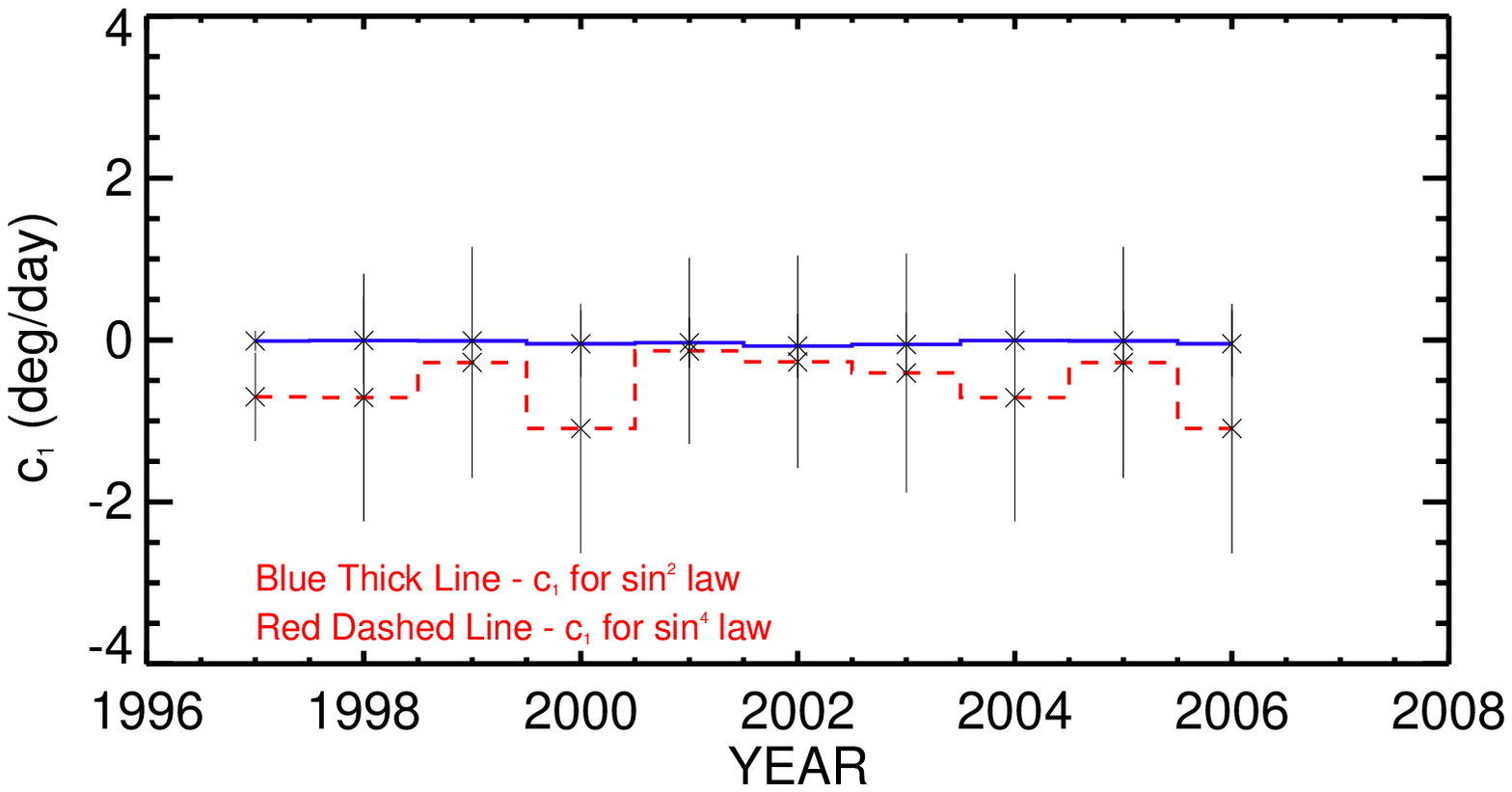}}\\
\end{tabular}
    \caption{For the years, 1997-2006, variation of coefficients $C_{0}$ and $C_{1}$ that
are obtained from the least square fit with a law up to $sin^{4} \theta$.}
 \end{center}
\end{figure}

\begin{figure}
\begin{center}
\vskip -6ex
      {\includegraphics[width=20pc,height=20pc]{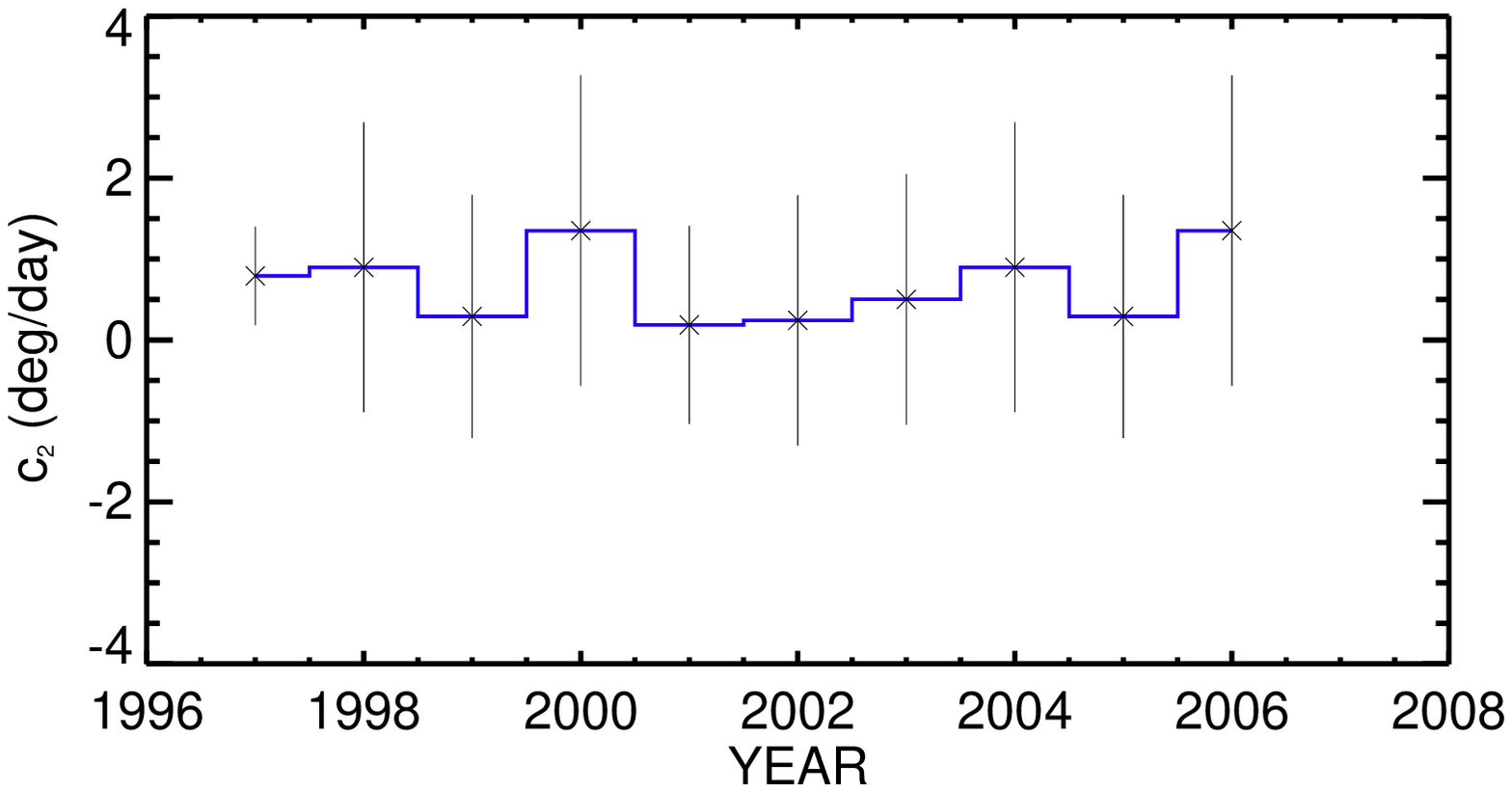}}
    \caption{For the years, 1997-2006, variation of coefficients $C_{2}$  that
is obtained from the least square fit with a law up to $sin^{4} \theta$.}
 \end{center}
\end{figure}

\subsection{Yearly variation of rotation rate of the coronal holes}
Many studies show that equatorial and mid-latitude rotation rates derived from sunspots and other
activity features of the photosphere vary with the activity of the solar cycle.
It is also interesting to examine whether, for all the latitudes considered
in this study, rotation rate of the coronal holes remain constant or varies
with the solar activity cycle . This study is not the first investigation.
By analyzing full disk spectroheliograms in HeI 10830 \AA, (Harvey and Sheeley 1987)
found that coronal holes rotate more rigidly during the low sunspot activity
and there is a substantial variation of rigidity during other periods of the solar activity.
For the years 1978-1986, by analyzing solar geophysical coronal hole data, Obridko and Shelting (1989)
came to a conclusion that 2-3 yrs before the solar maximum, coronal holes rotate
rigidly, whereas during rest of years of solar cycle, coronal holes rotate differentially.
 Navarro and Sanchez (1994) showed that coronal holes rotate slower during
maximum and vice verse during the minimum of the solar activity.

For different years, Figures 7-10 illustrate the latitudinal variation of rotation
rates of the coronal holes. As for the gaps in the data during June-October 1998 and January-February 1999,
we have considered 7 months (Jan-May, Nov-Dec) data while computing
the average rotation rates and 10 months (March-Dec) data during 1999. 
In all the Figures 7-10, left panel represents the latitudinal
variation of rotation rate which is fitted with a law up to $sin^{2} \theta$ and,
right panel is fitted with a law up to $sin^{4} \theta$. One can notice from all the
Figures 7-10 that, on average, for all the latitudes, rotation rate
of the coronal holes is independent of solar activity. Figures 11(a) and 11(b) illustrate the yearly variation
of coefficients $C_{0}$ and $C_{1}$ obtained from the least square fit with a law up to $sin^{4} \theta$.
Whereas Fig 12 illustrates the annual variation of coefficient $C_{2}$.
In all the Figures, although error bars appear to be large, all the three coefficients
$C_{0}$ (that represents equatorial rotation rate), $C_{1}$ (that represents
 rotation rate of the mid-latitudes) and $C_{2}$ ( that represents higher latitudes
and near polar regions) of least square fit with a law up to $sin^{4} \theta$
remain almost constant. Ultimately, from all these results, {\em it is inevitable to agree
that, for all the years and for all the latitudes, rotation rate of the coronal
holes is independent of solar activity}. Probably main reason for this yearly
constancy of rotation rate of coronal holes, which is different than the previous
studies, is that projectional effects for both the higher latitudes and longitudes
(from the central meridian) are taken into account in this study, whereas majority of previous
studies neglected the same.

\section{CONCLUSIONS}
For the years 1997-2006, SOHO UV 195 $\AA$ data is considered, coronal holes are detected,
their average heliographic coordinates such as latitude and longitudes are reasonably and accurately
are estimated. After taking into account the projectional effects for both latitude
and longitudes, rotation rates of the coronal holes are estimated. It is found that,
 for all the areas and irrespective of latitude, coronal holes rotate rigidly. Magnitudes of 
rotation rate of coronal holes are estimated to be $\sim$ $13.051 \pm 0.206$ deg/day for the equator,
 $12.993 \pm 0.064$ deg/day in the region of higher latitudes and, $12.999 \pm 0.329$ deg/day
near the polar regions. Yearly variation of rotation rate of coronal holes is examined. For all the years 1997-2006, 
coronal holes rotate rigidly and magnitude of equatorial, high latitude and polar region
rotation rates are independent of magnitude of solar activity.

\vskip 0.5cm
\centerline{\bf Acknowledgments}
\vskip 0.2cm
This work has been carried out under  ``CAWSES India Phase-II
program of Theme 1'' sponsored by Indian Space Research Organization(ISRO),
Government of India. SOHO is a mission of international cooperation between ESA and NASA.
\vskip 0.5cm

\centerline {\bf REFERENCES}
\vskip 0.5cm

Altschuler, M.D., Trotter, D.E., Orrall, F.Q.: 1972, Coronal Holes. Solar Phys. 26, 354.

 Antia, H. M., Basu, S \& Chitre, S. M. 1998, MNRAS, 298, 543

 Antia, H. M. \& Basu, S. 2010, ApJ, 720, 494

 Bagashvili, S. R.; Shergelashvili, B. M.; Japaridze, D. R.; Chargeishvili, B. B.; Kosovichev, A. G.; Kukhianidze, V.; Ramishvili, G.; Zaqarashvili, T. V.; Poedts, S.; Khodachenko, M. L.; De Causmaecker, P., 2017, Astrophys Astron, 603, 8

 Balthasar, H., Vazquez, M., \& Woehl, H. 1986 Astrophys Astron, 155, 87

 Bohlin, J. D. 1977, Solar Phys, 51, 377

Cranmer, S.R. 2009, Living Reviews in Solar Phys, 6, 3

Dalsgaard, C. J., \& Schou, J. 1988, in "Seismology of the Sun and Sun-like stars", p.149

Delaboudiniere, J. P. et al. 1995, Solar Phys, 162, 291

Freeland, S.L., Handy, B.N.: 1998, Solar Phys. 182, 497

Harvey, J. W., \& Sheeley, N. R., Jr. 1979, SSR, 23, 139

Harvey, J. W., \& Sheeley, N. R., Jr. 1987, Bulletin of the American Astronomical Society, Vol. 19, p.935

Hegde, M., Hiremath, K.M., Doddamani, V.H., Gurumath, S.R.: 2015, Astrophysics and Astronomy 36, 355.

Hiremath, K. M  2002, Astrophys Astron, 386, 674

Hiremath, K. M. 2009, Sun and Geosphere, 4, 16

Hiremath, K. M. \& Hegde, M.. 2013, ApJ, 763, 12

Howard, R., \& Harvey, J. 1970, Solar Phys, 12, 23

Howard, R., Gilman, P. I., \& Gilman, P. A. 1984, ApJ, 283, 373

Howe, R. 2009, Living Review in Solar Phys, 6, 1

Insley, J.E., Moore, V., \& Harrison, R.A. 1995, Solar Phys, 160, 1

Javaraiah, J. 2003, Solar Phys, 212, 23

Japaridze, D. R.; Bagashvili, S. R.; Shergelasvili, B. M.; Chargeishvili, B. B., 2015, Astrophys, 58, 575

Komm, R. W., Howard, R. F., \& Harvey, J. W. 1993, Solar Phys, 145, 1

Krieger, A.S., Timothy, A.F., \& Roelof, E.C. 1973, Solar Phys. 29, 505

Krista, L.D., Gallagher, P.T.: 2009, Solar Phys. 256, 87

Madjarska, M. S., \& Wiegelmann, T. 2009, Astrophys Astron, 503, 991

Navarro-Peralta, P., Sanchez-Ibarra, A.: 1994, Solar Phys. 153, 169

Neupert, W.M., \& Pizzo, V. 1974, J. Geophys. Res. 79, 3701

Newton, H. W., \& Nunn, M. L. 1951, MNRAS, 111, 413

Nolte, J. T., Krieger, A. S., Timothy, A. F., Gold, R. E., Roelof, E. C., Vaiana, G., Lazarus, A. J., Sullivan, J. D., \&  McIntosh, P. S. 1976, 46, 303

Oghrapishvili, N. B.; Bagashvili, S. R.; Maghradze, D. A.; Gachechiladze, T. Z.; Japaridze, D. R.; Shergelashvili, B. M.; Mdzinarishvili, T. G.; Chargeishvili, B. B., 2018, Advances in Space Research, 61, 3039

Prabhu, K.; Ravindra, B.; Hegde, Manjunath; Doddamani, Vijayakumar H., 2018, Astrophysics and Space Science, 363, 11

Obridko, V. N., \& Shelting, B. D. 1989, Solar Phys, 124, 73

Ro\v{s}a, D., Braj\v{s}a, R., Vr\v{s}nak, B., \&  W\"{o}hl, H. 1995, Solar Phys, 159, 393

Shelke, R. N., \& Pande, M. C. 1985, Solar Phys, 95, 193

Shugai, Yu. S., Veselovsky, I. S., \& Trichtchenko, L. D. 2009, Ge\&Ae, 49, 415

Sivaraman, K.R., Gupta, S.S., Howard, R.F.: 1993, Solar Phys. 146, 27

Skokić, I.; Brajša, R.; Roša, D.; Hržina, D.; Wöhl, H. 2014, Solar Physics, 289, 1471

Snodgrass, H. B. 1983, ApJ, 270, 288

Snodgrass, H. B., \& Ulrich, R. K. 1990, ApJ, 351, 309 

Soon, W., Baliunas, S., Posmentier, E. S., \&  Okeke, P. 2000, New Astronomy, 4, 563

Thompson, M. J. et al. 1996, Science, 272, 1300

Temmer, M., Vrsnak, B., Veronig, A.M.: 2007, Solar Phys. 241, 371

Timothy, A.F., Krieger, A.S., Vaiana, G.S.: 1975, Solar Phys. 42, 135

Thompson, M. J., Christensen-Dalsgaard, JÃ¸rgen, Miesch, Mark S. \&
Toomre, J. 2003, AAPR, 41, 599

Ulrich, R. K., Boyden, J. E., Webster, L., \& Shieber, T. 1988, in ESA,
Seismology of the Sun and Sun-Like Stars, 286, 325

Verbanac, G., Vr\v{s}nak, B., Veronig, A., \& Temmer, M. 2011, Astron Astrophys, 526, 20

Wagner, W.J.: 1975, Solar rotation as marked by extreme-ultraviolet coronal holes. Astrophys.
    J. Lett. 198, L141. DOI. ADS. 

Wagner, W.J.: 1976, Rotational Characteristics of Coronal Holes. In: Bumba, V., Kleczek, J.
    (eds.) Basic Mechanisms of Solar Activity, IAU Symposium 71, 41.

Wang, Y.M., Space Sci Rev, 2009, 144, 383

Wilcox, J. M., \& Howard, Robert 1970, Solar Phys, 13, 251

Wittmann, A. D. 1996, Solar Phys, 168, 211

Zirker, J. B. 1977, Reviews of Geophysics and Space Physics, 15, 257

Zurbuchen, Th.; Bochsler, P.; von Steiger, R. 1996, in Proceedings of the eigth international solar wind conference: Solar wind eight. AIP Conference Proceedings, 382, p. 273

\end{document}